\RequirePackage{ifpdf}
\documentclass[12pt]{JHEP3}

\usepackage{graphicx}
\usepackage{amsmath}
\usepackage{longtable}
\usepackage{caption}
\newcommand{\be}{\begin{eqnarray}}
\newcommand{\ee}{\end{eqnarray}}
\newcommand{\nn}{\nonumber}
\newcommand{\bn}{\begin{enumerate}}
\newcommand{\en}{\end{enumerate}}

\def\a{\alpha}

\def\Tr{{\rm Tr}}
\def\tr{{\rm tr}}

\newcommand{\RR}{{\mathbb R}}

\newcommand{\cN}{{\mathcal N}}

\newcommand{\Hom}{{\rm Hom}}
\newcommand{\cW}{{\mathcal W}}

\title{Factorization of the 3d superconformal index}

\author{ Chiung Hwang $^{1}$, Hee-Cheol Kim $^{2}$,  Jaemo Park$^{1,3}$

\\

$^1$Department of Physics, POSTECH, Pohang 790-784, Korea
\\
$^2$ School of Physics, Korea Institute for Advanced Study, Seoul 130-722, Korea.
\\
$^3$Postech Center for Theoretical Physics (PCTP), Postech, Pohang
  790-784, Korea

\\
\\
E-mail: \email{c\_hwang@postech.ac.kr, heecheol1@gmail.ac.kr,
jaemo@postech.ac.kr} } %%
\abstract{ We prove that 3d superconformal index for general
$\mathcal N=2$ $U(N)$  gauge group
with fundamentals and anti-fundmentals with/without Chern-Simons terms is factorized into
vortex and anti-vortex partition function. We show that for simple cases, 3d  vortex partition function
coincides with a suitable topological open string partition function. We provide much more elegant derivation
at the index level for $\mathcal N=2$ Seiberg-like dualities of unitary gauge groups with fundamantal matters and  $\mathcal N=4$ mirror symmetry}

\begin{document}

\section{Introduction}

Recently, there has been renewed interest in nonperturbative dualities between
three dimensional theories such as mirror symmetry and Seiberg-like
dualities. This is explained in part by the availability of
sophisticated tools such as the partition function on $S^3$ and the
superconformal index. Using
these tools, one can give impressive evidence
for various 3d dualities. Some of works in this area are \cite{Giveon09}-\cite{Aharony11}.

It turns out that the partition function has another interesting
property, i.e., it is factorized into vortex and anti-vortex
partition function \cite{Pasquetti:2011fj}. Schematically
\begin{equation}
Z(z,\bar{z})= Z_{vortex} (z) Z_{antivortex}(\bar{z})= |Z_{vortex} (z)|^2
\end{equation}
where $z$ traces the vortex number while $\bar{z}$ traces the anti-vortex number.
This is reminiscent of the  conformal blocks of the 2-dimensional
conformal field theories. The above factorization was shown to hold
for abelian gauge theories. Thus it is more desirable to show this
factorization for the general nonabelian cases. And it would be an
interesting question that the similar holds for 3d superconformal
index. In fact, it is recently shown that similar factorization
holds for 2-dimensional $\mathcal N=2$ supersymmetric partition function in
terms of vortex and anti-vortex partition function\cite{sungjay12, benini12}. Since 3d index
is the partition function defined on $S^1 \times S^2$, the two sphere partition function
is recovered from the 3d index by taking the radius of $S^1$ to be
small. Thus we expect that the factorization should hold for 3d
superconformal index as well.

The purpose of this paper is to show explicitly that such
factorization indeed occurs for 3d superconformal index. More
explicitly we show that for $U(N)$ gauge theories with $N_f$
fundamental and $\tilde{N}_f$ fundamentals and show that the index
is factorized into vortex and anti-vortex partition function on $R^2
\times S^1$ whenever $\max(N_f,\tilde{N}_f)\geq N$. This condition is
the condition of the existence of the vortex solutions of the
underlying field theories. This is done by explicit residue
evaluation of the associated matrix integral of the index, similar
to 2d case.

The factorized form of the index has a number of merits and we just
explore a few of them in this paper, relegating the full explorations
elsewhere. The first one is that we have the explicit expressions of
the index after the matrix integral. Obviously since we have the
explicit expressions for the index, it would be much more convenient
to explore the various dualities. Previously the index is expanded
in power series of the conformal dimension of the gauge invariant
BPS operators. In this way, one can check various dualities by
working out the index of the both sides to some orders in operator
dimensions. Though it certainly gives impressive evidences, in this
way the full analytic proof cannot be achieved. We will show that
explicit factorized formulae of the index reveal much more
transparent structures of the dualities. We will
see this by working out the index of the dual pairs of Aharony
duality with unitary gauge group. The proof of the equality of the
index is reduced to show the nontrivial identity of the
combinatorical character.

Furthermore in 2d case, the vortex partition function has the direct
connection to the topological open string amplitude. We expect that
similar holds for 3d vortex partition function since 2d vortex
partition function is so called the homological limit of 3d vortex
partition function. We show that vortex partition function is the
same as topological open string partition function for simple cases
but certainly has the obvious generalizations for much more numerous
examples, This is also resonant with the recent proposal by Iqbal
and Vafa \cite{Iqbal12} that the integrand of the  3d superconformal index is given
by the square of the topolgical open string amplitude. It would be
interesting to explore the precise relation between the 3d vortex
partition function and the open topological string.

 The content of
the paper is as follows. In section 2, we summarize the basic
structures of the superconformal index. We carefully study the
$U(1)$ gauge theory with a fundamental chiral multiplet with
Chern-Simons (CS) level $-1/2$ following \cite{DGG}, find
subtleties such as the relative phase of the different monopole
sector, in the usual index computation, which will be useful for
later computation. In section 3, we firstly work out the
factorization for $U(1)$ gauge theory without CS terms, which is
technically simpler. Then we summarize the factorization of the
general cases, deferring the full proof to the appendix.  We also
work out the explicit examples of the factorization and show the
associated vortex partition function admits topological open string
interpretation. Furthermore we show that in some of the examples  vortex
partition function  can be understood as 3d defect of the 5d field
theory. In section 4, we apply the factorized index to understand
the $\mathcal N=2$ Seiberg-like dualities for unitary gauge group, known as Aharony
duality. Factorized index reveals much more clearly such duality
should hold at the index level. We briefly touch upon the $\mathcal N=4$
Seiberg-like dualities and mirror symmetry and postpone the further
explorations elsewhere.

As this work is close to end, we receive the related paper by \cite{Dimofte12}.
As far as we understand , they do not give the general formulae for the
factorized index as we do.

\section{3d superconformal index}

\subsection{Summary of the 3d superconformal index}

Let us discuss the superconformal index for  $\cN=2$ $d=3$
superconformal field theories (SCFT). The bosonic subgroup of the
3d $\cN=2$ superconformal group  is $SO(2,3) \times SO(2) $. There
are three Cartan elements denoted by $\epsilon, j_3$ and $R$ which
come from three factors $SO(2)_\epsilon \times SO(3)_{j_3}\times
SO(2)_R $ in the bosonoic subgroup, respectively. The superconformal
index for an $\cN=2$ $d=3$ SCFT is defined as follows
\cite{Bhattacharya09}:
\begin{equation}
I(x,t)=\Tr (-1)^F \exp (-\beta'\{Q, S\}) x^{\epsilon+j_3}\prod_a
t_a^{F_a} \label{def:index}
\end{equation}
where $Q$ is a   supercharge with quantum numbers $\epsilon =
\frac{1}2, j_3 = -\frac{1}{2}$ and $R=1$, and $S= Q^\dagger$.  The
trace is taken over the Hilbert space in the SCFT on
$\mathbb{R}\times S^2$ (or equivalently over the space of local
gauge-invariant operators on $\RR^3$). The operators $S$ and $Q$
satisfy the following anti-commutation relation:
\begin{equation}
 \{Q, S\}=\epsilon-R-j_3 : = \Delta.
\end{equation}
As usual, only BPS states satisfying the bound $\Delta =0 $
contribute to the index, and therefore the index is independent of
the parameter $\beta'$. If we have additional conserved charges
$f_a$ commuting with the chosen supercharges ($Q,S$), we can turn on
the associated chemical potentials $t_a$, and then the index counts
the  number of BPS states weighted by their quantum
numbers.

The superconformal index is exactly calculable using the
localization technique \cite{Kim09,Imamura11}.  It can be written in
the following form:
\begin{eqnarray}\label{index}
& &I(x,t)= \nonumber\\
& &\sum_{m\in \mathbb{Z}} \int da\, \frac{1}{|\cW_m|}
e^{-S^{(0)}_{CS}(a,m)}e^{ib_0(a,m)} \prod_a t_{a}^{q_{0a}(m)}
x^{\epsilon_0(m)}\exp\left[\sum^\infty_{n=1}\frac{1}{n}f_{tot}(e^{ina},
t^n,x^n)\right].\qquad
\end{eqnarray}

The origin of this formula is as follows.
 To compute the trace over the Hilbert space on $S^2\times\RR$, we use path-integral on $S^2\times S^1$ with
 suitable boundary conditions
on the fields. The path-integral is evaluated using localization, which means that we have to sum or integrate
over all BPS saddle points. The saddle points are spherically symmetric configurations on $S^2\times S^1$ which
are labeled by magnetic fluxes on $S^2$ and holonomy along $S^1$. The magnetic fluxes are denoted by  $\{m\}$ and take
values in the cocharacter lattice of $G$ (i.e. in $\Hom(U(1),T)$, where $T$ is the maximal torus of $G$), while the eigenvalues of the holonomy are denoted $\{ a
\}$ and take values in $T$. $S_{CS}^{(0)}(a,m)$ is the classical action for the
(monopole+holonomy) configuration on $S^2\times S^1$, $\epsilon_0(m)$
is the Casmir energy of the vacuum state on $S^2$ with magnetic flux $m$, $q_{0a}(m)$ is the $f_a$-charge of the
vacuum state, and $b_0(a,m)$ represents the contribution coming from the electric charge of the vacuum state. The last factor comes from taking the trace over a Fock space built on a particular vacuum state. $|\cW_m|$ is the order of the Weyl group of the part of $G$ which is left unbroken by the magnetic fluxes $m$ . These ingredients in the formula for the index are given by the following explicit expressions:
\begin{eqnarray}\label{components}
&&S^{(0)}_{CS}(a,m) = i \sum_{\rho\in R_{F}} k \rho(m) \rho(a) , \\
&&b_0(a,m)=-\frac{1}{2}\sum_\Phi\sum_{\rho\in R_\Phi}|\rho(m)|\rho(a),\nonumber\\
&&q_{0a}(m) = -\frac{1}{2} \sum_\Phi \sum_{\rho\in R_\Phi} |\rho(m)| f_a (\Phi), \nonumber \\
&& \epsilon_0(m) = \frac{1}{2} \sum_\Phi (1-\Delta_\Phi) \sum_{\rho\in
R_\Phi} |\rho(m)|
- \frac{1}{2} \sum_{\alpha \in G} |\alpha(m)|, \nonumber
\end{eqnarray}
\begin{eqnarray}
&& f_{tot}(x,t,e^{ia})=f_{vector}(x,e^{ia})+f_{chiral}(x,t,e^{ia}),\nonumber\\
&& f_{vector}(x,e^{ia})=-\sum_{\alpha\in G} e^{i\alpha(a)} x^{|\alpha(m)|},\nonumber \\
&& f_{chiral}(x,t,e^{ia}) = \sum_\Phi \sum_{\rho\in R_\Phi} \left[
e^{i\rho(a)} \prod_a t_{a}^{f_{a}}
\frac{x^{|\rho(m)|+\Delta_\Phi}}{1-x^2}  -  e^{-i\rho(a)} \prod_a
t_{a}^{-f_{a}} \frac{x^{|\rho(m)|+2-\Delta_\Phi}}{1-x^2}\nonumber
\right]\label{universal}
\end{eqnarray}
where $\sum_{\rho\in R_F}, \sum_\Phi$, $\sum_{\rho\in R_\Phi}$ and $\sum_{\alpha\in G}$
represent summations over all fundamental weights of $G$, all chiral multiplets, all weights of the representation $R_\Phi$, and
all roots of $G$, respectively.

We will find it convenient to rewrite the integrand in \eqref{index} as a product of contributions
from the different multiplets.  First, note that the single particle index $f$ enters via the so-called Plethystic exponential:
\begin{align} \exp \bigg( \sum_{n=1}^\infty \frac{1}{n} f ( x^n, {t}^n,{z}^n=e^{ina}, m) \bigg) \end{align}
while we define $z_j=e^{ia_j}$.
It will be convenient to rewrite this using the $q$-product, defined for $n$ finite or infinite:
\begin{align} (z;q)_n = \prod_{j=0}^{n-1} (1 - z q^j). \end{align}
Specifically, consider a single chiral field $\Phi$, whose single
particle index is given by\footnote{Note that $a$ in $\rho(a)$ and the subscript $a$ in $t_a$ or $f_a$ denote
the different objects.} :
\begin{align} \sum_{\rho \in R_\Phi} \bigg( e^{i \rho(a)} {t_a}^{f_a(\Phi)} \frac{x^{ |\rho(m)|
+ \Delta_\Phi}}{1-x^2} - e^{-i \rho(a)} {t_a}^{-f_a(\Phi)} \frac{x^{
|\rho(m)| + 2-\Delta_\Phi}}{1-x^2} \bigg). \end{align}
Then we can
write the Plethystic exponential of this as follows:
\begin{align} \prod_{\rho \in R_\Phi} \exp \bigg[ \sum_{n=1}^\infty \frac{1}{n} \bigg( e^{i n\rho(a)} {t_a}^{n f_a(\Phi)}
\frac{x^{n |\rho(m)|  + n \Delta_\Phi}}{1-x^{2n}} - e^{-i n \rho(a)}
{t_a}^{-n f_a(\Phi)} \frac{x^{ n |\rho(m) | + 2n - n
\Delta_\Phi}}{1-x^{2n}} \bigg) \bigg]. \end{align}
By rewriting the
denominator as a geometric series and interchanging the order of
summations, one finds that this becomes:
\begin{align} \prod_{\rho \in R_\Phi} \frac{(e^{-i \rho(a)} {t_a}^{-f_a(\Phi)} x^{ |\rho(m)|
+ 2 -\Delta_\Phi} ; x^2)_\infty}{(e^{i \rho(a)} {t_a}^{f_a(\Phi)}
x^{ |\rho(m)| + \Delta_\Phi} ; x^2)_\infty}. \end{align}

The full index will involve a product of such factors over all the chiral fields in the theory, as well as the
contribution from the gauge multiplet.  It is given by:
\begin{align} I(x,t) = \sum_{m\in\mathbb{Z}} \oint \prod_j \frac{dz_j}{2 \pi i z_j} \frac{1}{|\mathcal W_m|} e^{-S_{CS}(m,a)} Z_{gauge}(x,z,m)
\prod_\Phi Z_\Phi(x,t,z,m) \end{align}
where:
\begin{align*} Z_{gauge}(x,z=e^{ia},m) = \prod_{\alpha \in ad(G)} x^{-\frac{|\alpha(m)|}{2} } \bigg(1 - e^{i \alpha(a)}
x^{ |\alpha(m)|} \bigg),  \end{align*}
\begin{align*}
 Z_\Phi(x,t,z,m)
 \!=\!\! \prod_{\rho \in R_\Phi}  \!\!\bigg(\!\!x^{(1- \Delta_\Phi) }
e^{-i \rho(a)} \prod_a {t_a}^{-f_a(\Phi)} \!\!\bigg)^{|\rho(m)|/2}
\frac{(e^{-i \rho(a)} {t_a}^{-f_a(\Phi)} x^{ |\rho(m)| + 2
-\Delta_\Phi} ; x^2)_\infty}{(e^{i \rho(a)} {t_a}^{f_a(\Phi)} x^{
|\rho(m)| + \Delta_\Phi} ; x^2)_\infty}.
\end{align*}
Note that by shifting $t_a\rightarrow t_ax^{c_a}$, one can change the value of the R-charge
$\Delta_{\Phi}$. Hence $\Delta_{\Phi}$ remains the free parameter for generic cases.

We are mainly interested in this ordinary index and work out the
factorization. However two important generalizations are worthy of mention, which will be
useful in comparison with the result of \cite{DGG} in the following subsection. The first
one is the notion of the generalized index. When we turn on the
chemical potential $t_a$, this can be regarded as turning on a Wilson
line for a fixed background gauge  field. The generalized index is
obtained when we turn on the nontrivial magnetic flux $n_a$ for the
corresponding background gauge field. Only the contribution to the
chiral multiplets are changed and this is given  by the replacement $\rho(m)\rightarrow \rho(m)+\sum_a f_a(\Phi)n_a$
\begin{eqnarray}
Z_\Phi(x,t,z,m)&=&\prod_{\rho \in R_\Phi}  \bigg(x^{(1- \Delta_\Phi) }
e^{-i \rho(a)} \prod_a {t_a}^{-f_a(\Phi)}
\bigg)^{|\rho(m)/2+\sum_a f_a(\Phi)n_a/2|}\nonumber\\
&&\qquad\frac{(e^{-i \rho(a)} {t_a}^{-f_a(\Phi)} x^{ |\rho(m)+\sum_a
f_a(\Phi)n_a| + 2 -\Delta_\Phi} ; x^2)_\infty}{(e^{i \rho(a)}
{t_a}^{f_a(\Phi)} x^{ |\rho(m)+\sum_a f_a(\Phi)n_a| + \Delta_\Phi} ;
x^2)_\infty}.
\end{eqnarray}
Here   $n_a$ should take integer value as does  $m_j$.

For every $U(N)$ gauge group, we can define another abelian symmetry
$U(1)_T$ whose conserved current is $*F$ of overall $U(1)$ factor.
To couple this topological current to background gauge field we
introduce $BF$ term $\int A_{BG} \wedge \tr dA+ \cdots$ and terms needed
for supersymmetric
completion. This introduces to the index
\begin{equation}
z^nw^{\sum_j m_j}
\end{equation}
where $n$ is the new discrete parameter representing the topological charge of $U(1)_T$ while $w$ is the chemical
potential for $U(1)_T$.

\subsection{Comparision to DGG}\label{DGGcomparison}
In the paper by Dimofte, Gaiotto and Gukov \cite{DGG} (DGG),
the simplet mirror pair of $\mathcal N=2$ theory was considered and
along with it revealed some subtleties in the index computation.
The claim is that the theory of one free chiral multiplet with global $U(1)$ symmetry at CS
level $\frac{1}{2}$ is mirror to  $U(1)$ gauge theory at CS level
$-\frac{1}{2}$, coupled to a single fundamental chiral multiplet.\footnote{We use a convention of the opposite sign for the CS level to DGG.} According to DGG, for
the free chiral theory the index is given by
\begin{equation}
\mathcal
I_\Delta(m;q,\zeta)=\left(-q^\frac{1}{2}\right)^{\frac{1}{2}(m+|m|)}\zeta^{-\frac{1}{2}(m+|m|)}
\prod_{r=0}^\infty\frac{1-q^{r+\frac{1}{2}|m|+1}\zeta^{-1}}{1-q^{r+\frac{1}{2}|m|}\zeta}.
\end{equation}
Note that we use the zero $R$-charge for the free chiral but
value of $R$-charge can be altered by shifting $\zeta \rightarrow
\zeta x^{\alpha}$ for a suitable $\alpha$. The index of  $U(1)$
theory is \cite{DGG}
\begin{equation}
\mathcal I_{U(1)}\left(m';q,\zeta'\right)=\sum_{m\in\mathbb
Z}\oint\frac{d\zeta}{2\pi i
\zeta}\zeta'^m\zeta^{m'}\left(-q^\frac{1}{2}\right)^{-\frac{1}{2}(m-|m|)}\zeta^{\frac{1}{2}(m-|m|)}
\prod_{r=0}^\infty\frac{1-q^{r+\frac{1}{2}|m|+1}\zeta^{-1}}{1-q^{r+\frac{1}{2}|m|}\zeta}.
\end{equation}
 It is  proved that $\mathcal
I_\Delta(m;q,\zeta)=\mathcal I_{U(1)}(m;q,\zeta)$.

In order to compare it to our index, let us slightly change the
variables as follows:
\begin{equation}
 I_{U(1)}\left(m';x^2,w\right)=\sum_{m\in\mathbb
Z}\oint\frac{dz}{2\pi i z}w^m
z^{m'}(-x)^{-\frac{1}{2}(m-|m|)}z^{\frac{1}{2}(m-|m|)}\prod_{k=0}^\infty\frac{1-z^{-1}x^{|m|+2+2k}}{1-zx^{|m|+2k}}. \label{DGGeqn}
\end{equation}
Note that $U(1)$ gauge theory has topological $U(1)$ global symmetry
whose current is given by $*F$ and $w$ corresponds to its chemical
potential. Under the mirror map, the global symmetry of chiral
theory is mapped to the topological symmetry. Hence $\zeta$ is mappd
to $w$. The expression appearing at DGG is slightly different from
the standard expression one obtains following the prescription specified at the previous
subsection or at \cite{Imamura11}. For $U(1)$ with CS level
$-1/2$,
  the index is given by\footnote{The factor $(-1)^{\frac{1}{2}(m-|m|)}$ will be explained in the next paragraph.}
\begin{equation}
I(x,w,m')=\sum_{m\in\mathbb Z}\oint\frac{dz}{2\pi i z}w^{m} z^{m'}
x^{|m|/2}(-z)^{\frac{1}{2}(m-|m|)}\prod_{k=0}^\infty\frac{1-z^{-1}x^{|m|+2+2k}}{1-z
x^{|m|+2k}}. \label{standard}
\end{equation}
The term  $x^{|m|/2}$ comes from the zero point energy
contribution. At first DGG expression appears to change the zero point energy
for positive and negative flux sector. However the  computation in \cite{Imamura11}
shows that the one-loop determinant is symmetric under $m\rightarrow
-m$ hence the zero point energy should be symmetric under
$m\rightarrow -m$, which comes from the suitable regularization of
one-loop determinant. The resolution is that if we assign
different $R$-charge in the free theory by $\zeta \rightarrow
\zeta q^{\alpha}$ we modify the $U(1)$ theory by $w\rightarrow
wx^{2\alpha}$. Using this freedom, if one shifts $w\rightarrow wx^{-1/2}$ one obtains
DGG eq. \eqref{DGGeqn} from the standard computation eq. \eqref{standard}.
On the other hand, in the $U(1)$ theory there's no freedom to
change the assigned $R$-charge of the charged chiral field and we assign
zero $R$-charge for the  scalar of the chiral multiplet. One might worry that
this $R$-charge can violate the unitarity of the SCFT.
However, the chiral field itself is not a gauge invariant operator.
Furthermore all of the gauge invariant operators of the theory are captured
by the index of the free chiral theory due to the mirror symmetry.
Thus the assigned zero $R$-charge does not lead to any inconsistency.
Furthermore one can show that the standard index of the $U(1)$ theory
eq. \eqref{standard} is equal to the free chiral theory with the canonical
$R$-charge 1/2, i.e.,
\begin{equation}
\mathcal
I_\Delta(m;q,\zeta)=\left(q^\frac{1}{2}\right)^{\frac{1}{4}(m+|m|)}(-\zeta)^{-\frac{1}{2}(m+|m|)}
\prod_{r=0}^\infty\frac{1-q^{r+\frac{1}{2}|m|+\frac{3}{4}}\zeta^{-1}}{1-q^{r+\frac{1}{2}|m|+\frac{1}{4}}\zeta}
\end{equation}
with $m=m'$, $q=x^2$ and $\zeta=w$. Thus if we use the standard
index computation we have the duality between $U(1)$ theory with CS
level $-1/2$  with one charged chiral with zero $R$-charge and the free
chiral with CS level 1/2 with the standard $R$-charge assignment.

On the other hand DGG assigns subtle relative phase factor $(-1)^{\frac{1}{2}(m+|m|)}$ between
positive and negative flux sector. This phase factor cannot be
derived from the usual index computation since it concerns on the
relative phase of the different flux sector. In DGG, this relative
phase factor have been checked extensively so we include this phase
in later computations. It turns out that this phase is crucial for
the factorization of the indices.

 For
reference, for $U(1)$ theory with CS level $\kappa$ with  $N_f$
fundamental chiral and  $\tilde N_f$ anti-fundamental chiral , the flavor symmetry is
$U(1)_A\times SU(N_f)\times SU(\tilde N_f)$. The index
we will use  is  as follows:
\begin{eqnarray}
I(x,t,\tilde{t},w,\kappa)&=&\sum_{m\in\mathbb Z}\oint\frac{dz}{2\pi i z}w^{m}x^{\frac{1}{2}(N_f+\tilde N_f)|m|}(-z)^{-\kappa m-\frac{1}{2}(N_f-\tilde N_f)|m|}\tau^{-\frac{1}{2}(N_f+\tilde N_f)|m|}\nonumber\\
&&\qquad\prod_{k=0}^\infty\left(\prod_{a=1}^{N_f}\frac{1-z^{-1}t_a^{-1}\tau^{-1}
x^{|m|+2+2k}}{1-z t_a \tau
x^{|m|+2k}}\right)\left(\prod_{a=1}^{\tilde N_f}\frac{1-z\tilde
t_a^{-1}\tau^{-1} x^{|m|+2+2k}}{1-z^{-1} \tilde t_a \tau
x^{|m|+2k}}\right)\nonumber\\
\end{eqnarray}
where $\tau, t_a, \tilde{t}_a$ are the fugacities for $U(1)_A$, Cartans of $SU(N_f), SU(\tilde{N}_f)$ respectively.
Note that we include the additional phase $(-1)^{-\kappa m-\frac{1}{2}(N_f-\tilde N_f)|m|}$ to the original index.
Similar factor will be included for non-abelian cases as well.

\section{Factorization
}
\subsection{Abelian without CS terms}
 We first consider the factorization for the abelian case
without CS terms. Similar but slightly more complicated derivation  works for $U(N)$ theory with
fundamentals and anti-fundamentals in the presence of Chern-Simons
terms. The general derivation is relegated to the appendix. The
superconformal index for $U(1)$ gauge theory is given by
\begin{equation}
I(x,t,w)=\sum_{m\in\mathbb Z}\oint\frac{dz}{2\pi i z}w^{m}\prod_\Phi Z_\Phi(x,t,z,m)
\end{equation}

If one considers $N_f$ fundamental and $\tilde N_f$ antifundamental chiral multiplets,
\begin{eqnarray}
&&\prod_\Phi Z_\Phi\left(x,t,\tilde{t},\tau,z,m\right)\nonumber\\
&=&x^{(1-\Delta_\Phi)(N_f+\tilde N_f)|m|/2}(-z)^{-(N_f-\tilde N_f)|m|/2}\tau^{-(N_f+\tilde N_f)|m|/2}\nonumber\\
&&\prod_{k=0}^\infty\left(\prod_{a=1}^{N_f}\frac{1-z^{-1}t_a^{-1}\tau^{-1}x^{|m|+2-\Delta_\Phi+2k}}{1-z t_a
\tau x^{|m|+\Delta_\Phi+2k}}\right)\left(\prod_{a=1}^{\tilde N_f}\frac{1-z \tilde t_a^{-1}\tau^{-1}x^{|m|+2
-\Delta_\Phi+2k}}{1-z^{-1}\tilde t_a\tau x^{|m|+\Delta_\Phi+2k}}\right)
\end{eqnarray}
where $t_a$ and $\tilde t_a$ correspond to fugacities for $SU(N_f)\times SU(\tilde N_f)$; $\tau$ is a
fugacity for $U(1)_A$ as in the previous section. We will set $\Delta_\Phi=0$, which can be restored by  deforming $\tau\rightarrow \tau x^{\Delta_\Phi}$.
Note that the infinite product in the expression only makes sense for $|x|<1$. Assuming $|t_a\tau|,|\tilde t_a\tau|<1$, which can be
extended by analytic continuation after integration, poles from the antifundamental part lie inside the integration contour.
In addition, the integrand also has a pole at the origin, which makes the integration difficult, for $N_f\geq\tilde N_f$.
Fortunately for $N_f>\tilde N_f$ one could change the integration variable $z\rightarrow 1/z$ to exclude the pole at
the origin and would take poles from the fundamental part, which are now inside the contour, instead of the poles
from the antifundamental part. For $N_f=\tilde N_f$ one should take account of the pole at the origin.

Firstly we deal with the $N_f>\tilde N_f$ case. Changing the variable $z\rightarrow 1/z$ is equivalent
to summing residues at poles outside the contour, which come from the fundamental
part: $z=t_{b}^{-1}\tau^{-1}x^{-|m|-2l}$ for $b=1,\cdots,N_f$ and $l=0,1,\cdots$. After performing the contour
integration the index is given by
\begin{eqnarray}
&&I^{N_f>\tilde N_f}(x,t,\tilde{t},\tau,w)\nonumber\\
&=&\sum_{m\in\mathbb Z}\sum_{b=1}^{N_f}\sum_{l=0}^\infty (-1)^{-(N_f-\tilde N_f)|m|/2}w^{m}t_{b}^{(N_f-\tilde N_f)|m|/2}
\tau^{-\tilde N_f|m|}x^{(N_f+\tilde N_f)|m|/2+(N_f-\tilde N_f)\left(|m|^2+2|m|l\right)/2}\nonumber\\
&&\left(\frac{\prod_{a=1}^{N_f}\prod_{k=0}^\infty1-t_{b} t_a^{-1}x^{2|m|+2l+2+2k}}
{\prod_{\substack{a=1,k=0\\((a,k)\neq(b,l))}}^{N_f,\infty}1-t_{b}^{-1} t_a x^{-2l+2k}}\right)
\left(\prod_{a=1}^{\tilde N_f}\prod_{k=0}^\infty\frac{1-t_{b}^{-1} \tilde t_a^{-1} \tau^{-2}
x^{-2l+2+2k}}{1-t_{b}\tilde t_a \tau^2 x^{2|m|+2l+2k}}\right).
\end{eqnarray}
Let us rewrite the terms $\prod_{\substack{a=1,k=0\\((a,k)\neq(b,l))}}^{N_f,\infty}1-t_{b}^{-1} t_a x^{-2l+2k}$ and
$\prod_{a=1}^{\tilde N_f}\prod_{k=0}^\infty1-t_{b}^{-1} \tilde t_a^{-1} \tau^{-2} x^{-2l+2+2k}$ as follows:
\begin{eqnarray}
&&\prod_{\substack{a=1,k=0\\((a,k)\neq(b,l))}}^{N_f,\infty}1-t_{b}^{-1} t_a x^{-2l+2k}\\
%&=&\left(\prod_{a=1}^{N_f}\prod_{k=0}^{l-1}1-t_{b}^{-1} t_a x^{-2l+2k}\right)\left(\prod_{\substack{a=1,k=0\\((a,k)\neq(b,0))}}^{N_f,\infty}1-t_{b}^{-1} t_a x^{2k}\right)\nonumber\\
&=&\left(\prod_{a=1}^{N_f}\prod_{k=0}^{l-1}-t_{b}^{-1} t_a x^{-2l+2k}\!\right)\left(\prod_{a=1}^{N_f}\prod_{k=0}^{l-1}1-t_{b} t_a^{-1} x^{2+2k}\right)\left(\prod_{\substack{a=1,k=0\\((a,k)\neq(b,0))}}^{N_f,\infty}1-t_{b}^{-1} t_a x^{2k}\!\right)\nonumber
\end{eqnarray}
and
\begin{eqnarray}
&&\prod_{a=1}^{\tilde N_f}\prod_{k=0}^\infty1-t_{b}^{-1} \tilde t_a^{-1} \tau^{-2} x^{-2l+2+2k}\\
%&=&\left(\prod_{a=1}^{\tilde N_f}\prod_{k=0}^{l-1}1-t_{b}^{-1} \tilde t_a^{-1} \tau^{-2} x^{-2l+2+2k}\right)\left(\prod_{a=1}^{\tilde N_f}\prod_{k=0}^\infty1-t_{b}^{-1} \tilde t_a^{-1} \tau^{-2} x^{2+2k}\right)\nonumber\\
\!&\!=\!&\!\left(\prod_{a=1}^{\tilde N_f}\prod_{k=0}^{l-1}-t_{b}^{-1} \tilde t_a^{-1} \tau^{-2} x^{-2l+2+2k}\!\right)\!\!\!\!
\left(\prod_{a=1}^{\tilde N_f}\prod_{k=0}^{l-1}1-t_{b} \tilde t_a \tau^2 x^{2k}\!\right)\!\!\!\!\left(\prod_{a=1}^{\tilde N_f}
\prod_{k=0}^\infty1-t_{b}^{-1} \tilde t_a^{-1} \tau^{-2} x^{2+2k}\!\right).\nonumber
\end{eqnarray}
Using these one can rewrite the index as follows:
\begin{eqnarray}
\hspace{-1.2cm}&&I^{N_f>\tilde N_f}(x,t,\tilde{t},\tau,w)\nonumber\\
\hspace{-1.2cm}\!&\!=\!&\!\sum_{m\in\mathbb Z}\sum_{b=1}^{N_f}\sum_{l=0}^\infty (-1)^{-(N_f-\tilde N_f)|m|/2}w^{m}t_{b}^{(N_f-\tilde N_f)|m|/2}
\tau^{-\tilde N_f|m|}x^{(N_f+\tilde N_f)|m|/2+(N_f-\tilde N_f)\left(|m|^2+2|m|l\right)/2}\nonumber\\
\hspace{-1.2cm}&&\qquad\times\frac{\left(\prod_{a=1}^{N_f}\prod_{k=0}^\infty1-t_{b} t_a^{-1}x^{2+2k}\right)/\left(\prod_{a=1}^{N_f}
\prod_{k=0}^{|m|+l-1}1-t_{b} t_a^{-1}x^{2+2k}\right)}{\left(\!\prod_{a=1}^{N_f}\prod_{k=0}^{l-1}-t_{b}^{-1}
t_a x^{-2l+2k}\!\right)\!\left(\!\prod_{a=1}^{N_f}\prod_{k=0}^{l-1}1-t_{b} t_a^{-1} x^{2+2k}\!\right)\!\left(\!\prod_{\substack{a=1,k=0\\
((a,k)\neq(b,0))}}^{N_f,\infty}1-t_{b}^{-1} t_a x^{2k}\!\right)}\nonumber\\
\hspace{-1.2cm}&&\qquad\times\frac{\left(\!\prod_{a=1}^{\tilde N_f}\prod_{k=0}^{l-1}\!-t_{b}^{-1} \tilde t_a^{-1} \tau^{-2} x^{-2l+2+2k}\!\right)\!\!
\left(\!\prod_{a=1}^{\tilde N_f}\prod_{k=0}^{l-1}1\!-\!t_{b} \tilde t_a \tau^2 x^{2k}\!\right)\!\!\left(\!\prod_{a=1}^{\tilde N_f}
\prod_{k=0}^\infty1\!-\!t_{b}^{-1} \tilde t_a^{-1} \tau^{-2} x^{2+2k}\!\right)}{\left(\prod_{a=1}^{\tilde N_f}
\prod_{k=0}^{\infty}1-t_{b} \tilde t_a \tau^2 x^{2k}\right)/\left(\prod_{a=1}^{\tilde N_f}\prod_{k=0}^{|m|+l-1}1-t_{b}
\tilde t_a \tau^2 x^{2k}\right)}\nonumber\\
\hspace{-1.2cm}\!&\!=\!&\!\sum_{m\in\mathbb Z}\sum_{b=1}^{N_f}\sum_{l=0}^\infty(\!-1\!)^{-(N_f\!-\!\tilde N_f)(|m|+2l)/2}w^{m}t_{b}^{(N_f\!-\!\tilde N_f)(|m|+2l)/2}
\tau^{\!-\!\tilde N_f(|m|+2l)}x^{(N_f\!+\!\tilde N_f)(|m|+2l)/2+(N_f\!-\!\tilde N_f)\left[(|m|+l)^2+l^2\right]/2}\nonumber\\
\hspace{-1.2cm}&&\qquad\times\left(\prod_{k=0}^\infty\frac{\prod_{a=1(\neq b)}^{N_f}1-t_{b} t_a^{-1}x^{2+2k}}{\prod_{a=1}^{\tilde N_f}1-t_{b}
\tilde t_a \tau^2 x^{2k}}\right)\left(\prod_{k=0}^\infty\frac{\prod_{a=1}^{\tilde N_f}1-t_{b}^{-1} \tilde t_a^{-1}
\tau^{-2} x^{2+2k}}{\prod_{a=1(\neq b)}^{N_f}1-t_{b}^{-1} t_a x^{2k}}\right)\nonumber\\
\hspace{-1.2cm}&&\qquad\times\left(\prod_{k=0}^{|m|+l-1}\frac{\prod_{a=1}^{\tilde N_f}1-t_{b} \tilde t_a \tau^2 x^{2k}}
{\prod_{a=1}^{N_f}1-t_{b} t_a^{-1}x^{2+2k}}\right)\left(\prod_{k=0}^{l-1}\frac{\prod_{a=1}^{\tilde N_f}1-t_{b}
\tilde t_a \tau^2 x^{2k}}{\prod_{a=1}^{N_f}1-t_{b} t_a^{-1} x^{2+2k}}\right).
\end{eqnarray}
In order to proceed further one should rearrange the summations. Thanks to the
symmetry $|m|+l\leftrightarrow l$ one can rearrange the summations as
$\sum_{m\in\mathbb Z}\sum_{l=0}^\infty=\sum_{n=0}^\infty\sum_{{\bar n}=0}^\infty$ where
 $n\equiv l+\frac{|m|}{2}+\frac{m}{2}$, $\bar n\equiv l+\frac{|m|}{2}-\frac{m}{2}$. Finally the index can be written in the factorized form:
\begin{eqnarray}
&&I^{N_f>\tilde N_f}(x,t,\tilde{t},\tau,w)\nonumber\\
\!&\!=\!&\!\sum_{b=1}^{N_f}\!\sum_{n=0}^\infty\!\sum_{{\bar n}=0}^\infty\! (-1)^{-(N_f\!-\!\tilde N_f)(n\!+\!\bar n)/2}w^{\sum(n\!-\!\bar n)}t_{b}^{(N_f\!-\!\tilde N_f)(n\!+\!\bar n)/2}
\tau^{-\tilde N_f(n\!+\!\bar n)}x^{(N_f\!+\!\tilde N_f)(n\!+\!\bar n)/2\!+\!(N_f\!-\!\tilde N_f)\left(n^2\!+\!\bar n^2\right)/2}\nonumber\\
&&\qquad\times\left(\prod_{k=0}^\infty\frac{\prod_{a=1(\neq b)}^{N_f}1-t_{b} t_a^{-1}x^{2+2k}}{\prod_{a=1}^{\tilde N_f}1-t_{b}
\tilde t_a \tau^2 x^{2k}}\right)\left(\prod_{k=0}^\infty\frac{\prod_{a=1}^{\tilde N_f}1-t_{b}^{-1} \tilde t_a^{-1}
\tau^{-2} x^{2+2k}}{\prod_{a=1(\neq b)}^{N_f}1-t_{b}^{-1} t_a x^{2k}}\right)\nonumber\\
&&\qquad\times\left(\prod_{k=0}^{n-1}\frac{\prod_{a=1}^{\tilde N_f}1-t_{b} \tilde t_a \tau^2 x^{2k}}{\prod_{a=1}^{N_f}1-t_{b}
t_a^{-1}x^{2+2k}}\right)\left(\prod_{k=0}^{\bar n-1}\frac{\prod_{a=1}^{\tilde N_f}1-t_{b} \tilde t_a \tau^2 x^{2k}}
{\prod_{a=1}^{N_f}1-t_{b} t_a^{-1} x^{2+2k}}\right)\nonumber\\
\!&\!=\!&\!\sum_{b=1}^{N_f}\left[\prod_{k=0}^\infty\left(\frac{\prod_{a=1(\neq b)}^{N_f}1-t_{b} t_a^{-1}x^{2+2k}}
{\prod_{a=1}^{\tilde N_f}1-t_{b} \tilde t_a \tau^2 x^{2k}}\right)\left(\frac{\prod_{a=1}^{\tilde N_f}1-t_{b}^{-1}
\tilde t_a^{-1} \tau^{-2} x^{2+2k}}{\prod_{a=1(\neq b)}^{N_f}1-t_{b}^{-1} t_a x^{2k}}\right)\right]\nonumber\\
&&\times\!\sum_{n=0}^\infty\!\!\left[\!(-1)^{-(N_f\!-\!\tilde N_f)n/2}w^{n}t_{b}^{(N_f\!-\!\tilde N_f)n/2}\tau^{-\tilde N_f n}x^{(N_f\!+\!\tilde N_f) n/2\!+\!(N_f\!-\!\tilde N_f)
n^2/2}\prod_{k=0}^{n-1}\frac{\prod_{a=1}^{\tilde N_f}1\!-\!t_{b} \tilde t_a \tau^2 x^{2k}}{\prod_{a=1}^{N_f}1-t_{b}
t_a^{-1}x^{2+2k}}\right]\nonumber\\
&&\times\!\sum_{{\bar n}=0}^\infty\!\!\left[\!(-1)^{-(N_f\!-\!\tilde N_f)\bar n/2}w^{-\bar n}t_{b}^{(N_f\!-\!\tilde N_f)\bar n/2}\tau^{-\tilde N_f\bar n}
x^{(N_f\!+\!\tilde N_f)\bar n/2\!+\!(N_f\!-\!\tilde N_f)\bar n^2/2}\prod_{k=0}^{\bar n-1}\frac{\prod_{a=1}^{\tilde N_f}1-t_{b}
\tilde t_a \tau^2 x^{2k}}{\prod_{a=1}^{N_f}1-t_{b} t_a^{-1} x^{2+2k}}\right]\nonumber\\
\end{eqnarray}
More concisely,
\begin{eqnarray}\label{expression 1}
&&I^{N_f>\tilde N_f}(x=e^{-\gamma},t=e^{iM},\tilde t=e^{i\tilde M},\tau=e^{i\mu},w)\nonumber \\
\!&\!=\!&\!\sum_{b=1}^{N_f}\left[\prod_{k=0}^\infty\left(\frac{\prod_{a=1(\neq b)}^{N_f}1-t_{b} t_a^{-1}x^{2+2k}}
{\prod_{a=1}^{\tilde N_f}1-t_{b} \tilde t_a \tau^2 x^{2k}}\right)\left(\frac{\prod_{a=1}^{\tilde N_f}1-t_{b}^{-1}
\tilde t_a^{-1} \tau^{-2} x^{2+2k}}{\prod_{a=1(\neq b)}^{N_f}1-t_{b}^{-1} t_a x^{2k}}\right)\right]\nonumber\\
&&\times\sum_{n=0}^\infty\left[(-1)^{-(N_f-\tilde N_f)n/2}(-w)^{n}\prod_{k=0}^{n-1}\frac{\prod_{a=1}^{\tilde N_f}2\sinh\frac{
-i\tilde M_a-iM_{b}-2i\mu+2\gamma k}{2}}{2\sinh\gamma(k-n)\prod_{a=1(\neq b)}^{N_f}2\sinh\frac{iM_a-iM_{b}+2\gamma(1+k)}{2}}\right]\nonumber\\
&&\times\sum_{{\bar n}=0}^\infty\left[(-1)^{-(N_f-\tilde N_f)\bar n/2}(-w)^{-\bar n}\prod_{k=0}^{\bar n-1}
\frac{\prod_{a=1}^{\tilde N_f}2\sinh\frac{-i\tilde M_a-iM_{b}-2i\mu+2\gamma k}{2}}
{2\sinh\gamma(k-\bar n)\prod_{a=1(\neq b)}^{N_f}2\sinh\frac{iM_a-iM_{b}+2\gamma(1+k)}{2}}\right]\label{u1}\nonumber\\
\end{eqnarray}
where we identified some parameters as follows: $x= e^{-\gamma}$, $t_a= e^{iM_a}$, $\tilde t_a= e^{i\tilde M_a}$ and $\tau= e^{i\mu}$. The second last and last lines correspond to the $\mathcal N=2$ vortex and antivortex partition
functions on $\mathbb R^2\times S^1$ as appearing in \cite{Pasquetti:2011fj, KKKL12}.
In \cite{KKKL12}, vortex quantum mechanics is considered and the vortex zero sector is not handled. Thus we cannot
 directly compare the prefactor corresponding to the one-loop determinant. Perturbative part was checked in the other
 example \cite{ Pasquetti:2011fj} by matching to the 2d result. One can easily check that prefactor  of eq. \eqref{u1} also factorizes and gives rise to the one-loop determinant which matches the known result.

The index for $N_f<\tilde N_f$ is simply obtained by interchanging $t_a\leftrightarrow\tilde t_a$:
\begin{eqnarray}
&&I^{N_f<\tilde N_f}(x=e^{-\gamma},t=e^{iM},\tilde t=e^{i\tilde M},\tau=e^{i\mu},w)\nonumber \\
\!&\!=\!&\!\sum_{b=1}^{\tilde N_f}\left[\prod_{k=0}^\infty\left(\frac{\prod_{a=1(\neq b)}^{\tilde N_f}1
-\tilde t_{b}\tilde t_a^{-1}x^{2+2k}}{\prod_{a=1}^{N_f}1-\tilde t_{b}t_a \tau^2 x^{2k}}\right)
\left(\frac{\prod_{a=1}^{N_f}1-\tilde t_{b}^{-1}t_a^{-1} \tau^{-2} x^{2+2k}}{\prod_{a=1(\neq b)}^{\tilde N_f}1
-\tilde t_{b}^{-1}\tilde t_a x^{2k}}\right)\right]\nonumber\\
&&\times\sum_{n=0}^\infty\left[(-1)^{-(\tilde N_f-N_f)n/2}(-w)^{n}\prod_{k=0}^{n-1}\frac{\prod_{a=1}^{N_f}2\sinh\frac{-iM_a-i\tilde M_{b}
-2i\mu+2\gamma k}{2}}{2\sinh\gamma(k-n)\prod_{a=1(\neq b)}^{\tilde N_f}2\sinh\frac{i\tilde M_a-i\tilde M_{b}+2\gamma(1+k)}{2}}\right]\nonumber\\
&&\times\sum_{\bar{n}=0}^\infty\left[(-1)^{-(\tilde N_f-N_f)\bar n/2}(-w)^{-\bar n}\prod_{k=0}^{\bar n-1}
\frac{\prod_{a=1}^{N_f}2\sinh\frac{-iM_a-i\tilde M_{b}-2i\mu+2\gamma k}{2}}{2\sinh\gamma(k-\bar n)\prod_{a=1}^{\tilde N_f}2\sinh\frac{i\tilde M_a-i\tilde M_{b}+2\gamma(1+k)}{2}}\right]\nonumber\\
\end{eqnarray}
For $N_f=\tilde N_f$ the integrand also has poles both at the origin and at the infinity. The residue at the origin is given by
\begin{eqnarray}
&&\textrm{Res}(\ldots,0)\nonumber\\
&=&x^{N_f|m|}\tau^{-N_f|m|}\lim_{z\rightarrow0}\prod_{a=1}^{N_f}\prod_{k=0}^\infty\frac{z-t_a^{-1}\tau^{-1}x^{|m|+2+2k}}{1-z t_a\tau x^{|m|+2k}}\frac{1-z \tilde t_a^{-1}\tau^{-1}x^{|m|+2+2k}}{z-\tilde t_a\tau x^{|m|+2k}}\nonumber\\
&=&x^{N_f|m|}\tau^{-N_f|m|}\prod_{a=1}^{N_f}\prod_{k=0}^{\infty}t_a^{-1}\tilde t_a^{-1}\tau^{-2}x^2\rightarrow0
\end{eqnarray}
assuming $|t_a^{-1}\tilde t_a^{-1}\tau^{-2}x^2|<1$, which doesn't spoil the original range of parameters that we already assumed. Since the residues at the other poles are the same as those for $N_f\neq\tilde N_f$, both results for $N_f>\tilde N_f$ and $N_f<\tilde N_f$ even work for $N_f=\tilde N_f$.
One can also check the result is reduced to the known result of 2d partition function of $\mathcal N=2$ theories
in the small radius limit of $S^1$ \cite{sungjay12, benini12}.
The same is true of non-abelian cases, which we will summarize in the next
subsection.

\subsection{Factorization: summary of nonabelian cases}
Now we summarize the factorized index formula for non-abelian cases
in the presence of CS terms.
The superconformal index in the presence of nonzero Chern-Simons
term is written as
\begin{equation}
I(x,t,w,\kappa)=\sum_{\vec m\in\mathbb Z^{N}/S_{N}}\oint\prod_j\frac{dz_j}{2\pi i z_j}\frac{1}{|\mathcal W_m|}w^{\sum_j m_j}e^{-S_{CS}(a,m)}Z_{gauge}(x,z,m)\prod_\Phi Z_\Phi(x,t,z,m)
\end{equation}
where $S^{(0)}_{CS}(a,m)=i\sum_{\rho\in R_F}\kappa
\rho(m)\rho(a)$. The Chern-Simons term with level $\kappa$ induces
the classical action term in the path integral. It leads to the pole
at $z_i=0$ or $z_i = \infty$ according to the sign of $\kappa m$. As shown in
the appendix,  one can show that the residues at these poles are
zero. The contour integral over the holonomy variables of the gauge
group can be written as \be
    \hspace{-1cm}&&I^{N_f,\tilde{N}_f}(x,t,\tilde{t},w,\kappa) \nonumber \\
    \hspace{-1cm}&\!=\!&\!\! \frac{1}{N!(N_f\!-\!N)!}\!\sum_{\sigma(t)}I^{pert}(x,\sigma(t),\tilde{t},\tau) \!\!\!\left[\sum_{\vec{n}=0}^\infty(-w)^{n}I_{\{n_j\}}(x,\sigma(t),\tilde{t},\tau,\kappa)\!\right]
    \!\!\!\left[\sum_{\vec{\bar{n}}=0}^\infty(-w)^{-\bar{n}}I_{\{\bar{n}_j\}}(x,\sigma(t),\tilde{t},\tau,-\kappa)\!\right]\nonumber\\
    \hspace{-1cm}
\ee
where $n=\sum_j n_j,\bar{n}=\sum_j\bar{n}_j$ and $\sigma(t)$
denotes the permutation of $t_a$'s. Here the perturbative and vortex
contributions are given by \be
    \hspace{-.8cm}I^{pert}(x,t,\tilde{t},\tau) \!&\!=\!&\! \prod_{i\neq j}^N2\sinh\frac{iM_i-iM_j}{2}
    \prod_{j=1}^N\prod_{k=0}^\infty\!\!\left[\!\prod_{a=1(\neq j)}^{N_f}\frac{1\!-\!t_j t_a^{-1}x^{2k+2}}{1\!-\!t^{-1}_j t_a x^{2k}}\right]\!\!\!\!\left[\prod_{a=1}^{\tilde{N}_f}\frac{1\!-\!t_j^{-1}\tilde{t}_a^{-1}\tau^{-2}x^{2k+2}}{1\!-\!t_j\tilde{t}_a
    \tau^2x^{2k}}\right], \nn \\
    \hspace{-.8cm}I_{\{n_j\}}(x,t,\tilde{t},\tau,\kappa)\! &\!=\!&\! (-1)^{-\kappa n-(N_f-\tilde N_f)n/2}e^{i\kappa\sum_j( M_j n_j+\mu n_j+i\gamma n_j^2)}\nonumber\\
    \hspace{-.8cm}&&\prod_{j=1}^N\prod_{k=1}^{n_j}\frac{\prod_{a=1}^{\tilde{N}_f}2\sinh\frac{-i\tilde{M}_a-iM_j-2i\mu+2\gamma (k-1)}{2}}
    {\prod_{i=1}^N2\sinh\frac{iM_i-iM_j+2\gamma(k-1-n_i)}{2}\prod_{a=N+1}^{N_f}2\sinh\frac{iM_a-iM_j+2\gamma k}{2}}\qquad
\ee where the prefactor depending on $\kappa$ in the vortex part
appears due to the nonzero Chern-Simons term.

\subsection{Factorization of mirror of one free chiral}
Let us consider a $U(1)$ theory with a single chiral multiplet in
the fundamental representation. If one also turns on the level
$-\frac{1}{2}$ CS interaction and the fixed background magnetic flux
$m'$ corresponding to the topological global symmetry $U(1)_T$, the
index is given by
\begin{equation}\label{a half CS0}
I(x,w',m')=\sum_{m\in\mathbb Z}\oint\frac{dz}{2\pi i z}w'^{m}z^{m'}
x^{\frac{1}{2}|m|}(-z)^{\frac{1}{2}(m-|m|)}\prod_{k=0}^\infty\frac{1-z^{-1}x^{|m|+2-\Delta_\phi+2k}}{1-z
x^{|m|+\Delta_\phi+2k}}.
\end{equation}

This  can be written as
\begin{equation}\label{a half CS}
I(x,w,m')=\sum_{m\in\mathbb Z}\oint\frac{dz}{2\pi i
z}w^{m}z^{m'}(-x)^{-\frac{1}{2}(m-|m|)}z^{\frac{1}{2}(m-|m|)}\prod_{k=0}^\infty\frac{1-z^{-1}x^{|m|+2+2k}}{1-z
x^{|m|+2k}}
\end{equation}
where we redefined $w'x^{\frac{1}{2}}\rightarrow
w$. A factor $x^\frac{1}{2}$ is additionally absorbed to $w$ for
later convenience.

One may consider its mirror description, a single free chiral theory
with the level $\frac{1}{2}$ CS interaction. As introduced in the previous section the index for the
mirror description is given by \cite{DGG}
\begin{equation}\label{DGGeqn2}
\mathcal
I_\Delta(m;q,\zeta)=\left(-q^\frac{1}{2}\right)^{\frac{1}{2}(m+|m|)}\zeta^{-\frac{1}{2}(m+|m|)}\prod_{r=0}^\infty\frac{1-q^{r+\frac{1}{2}|m|+1}\zeta^{-1}}{1-q^{r+\frac{1}{2}|m|}\zeta}
\end{equation}
where $m$ and $\zeta$ are magnetic flux and the Wilson line of the
fixed background $U(1)$ vector field coupling to the conserved
current of the $U(1)$ flavor symmetry. Again the parameters are identified
with ours as follows:
\begin{equation}
m=m',\qquad q=x^2,\qquad \zeta=w.
\end{equation}
 It was
argued in \cite{DGG} that the index \eqref{a half CS}
agrees with \eqref{DGGeqn2}.

Here we revisit the index agreement using the factorized form of the index. The
factorized form of \eqref{a half CS} is given as follows\footnote{Compared with
the general formula, the power in $x$ has $x^{-\frac{n(n+1)}{2}}$ while the general
formula appearing at the appendix has $x^{-\frac{n^2}{2}}$. The reason is that \eqref{a half CS}
matches with the free theory with zero $R$-charge for the free chiral while the standard factorized
formula matches with the free chiral with canonical $R$-charge. Two expressions are related by the shift
$w\rightarrow wx^{-\frac{1}{2}}$.} :
\begin{eqnarray}
\hspace{-.4cm}&&I(x=e^{-\gamma},w,m')\nonumber\\
\hspace{-.4cm}\!&\!=\!&\!\sum_{n=0}^\infty\!\!\left[\!w^{n}x^{-m'n-\frac{n(n+1)}{2}}\!\!\left(\prod_{k=1}^{n}\!2\sinh\gamma k\!\right)^{-1}\!\right]\!\!\!\times\!\!\sum_{{\bar n}=0}^\infty\!\!\left[\!(-w)^{-\bar n}x^{-m'\bar n+\frac{\bar n(\bar n+1)}{2}}\!\!\left(\prod_{k=1}^{\bar n}\!2\sinh\gamma k\!\right)^{-1}\!\right].\nonumber\\
\hspace{-.4cm}
\end{eqnarray}
As before the first summation corresponds to the vortex partition
function while the second summation corresponds to the antivortex
partition function. One may check that the vortex partition function
can  be written as a Plethystic exponential:
\begin{eqnarray}
Z_{vortex}(x,w,m')\!&\equiv&\!\sum_{n=0}^\infty\left[w^{n}x^{-m'n-\frac{n(n+1)}{2}}\left(\prod_{k=1}^{n}2\sinh\gamma k\right)^{-1}\right]\nonumber\\
&=&\exp\left[\sum_{n=1}^\infty\frac{1}{n}\frac{w^n x^{-m'n}}{1-x^{2n}}\right].
\end{eqnarray}
Likewise, the antivortex partition function also has the Plethystic
exponential form:
\begin{eqnarray}
Z_{anti}(x,w,m')\!&\!\equiv\!&\!\sum_{{\bar n}=0}^\infty\left[(-w)^{-\bar n}x^{-m'\bar n+\frac{\bar n(\bar n+1)}{2}}\left(\prod_{k=1}^{\bar n}2\sinh\gamma k\right)^{-1}\right]\nonumber\\
&=&\exp\left[-\sum_{n=1}^\infty\frac{1}{n}\frac{w^{-n}x^{n(-m'+2)}}{1-x^{2n}}\right].
\end{eqnarray}
On the other hand, it was pointed out in \cite{DGG} that
the free chiral index \eqref{DGGeqn} has a more concise form as
follows:
\begin{equation}
\mathcal
I_\Delta(m;q,\zeta)=\prod_{r=0}^\infty\frac{1-q^{r-\frac{1}{2}m+1}\zeta^{-1}}{1-q^{r-\frac{1}{2}m}\zeta}.
\end{equation}
One can see that the denominator, which comes from the scalar, is
exactly the vortex partition function while the numerator, which
comes from the fermion, is the antivortex partition function:
\begin{eqnarray}
Z_{vortex}(q^\frac{1}{2},\zeta,m)=\exp\left[\sum_{n=1}^\infty\frac{1}{n}\frac{\zeta^n q^{-\frac{1}{2}mn}}{1-q^{n}}\right]&=&\prod_{r=0}^\infty\frac{1}{1-q^{r-\frac{1}{2}m}\zeta},\\
Z_{anti}(q^\frac{1}{2},\zeta,m)=\exp\left[-\sum_{n=1}^\infty\frac{1}{n}\frac{\zeta^{-n}q^{(-\frac{1}{2}m+1)n}}{1-q^{n}}\right]&=&\prod_{r=1}^\infty
1-q^{r-\frac{1}{2}m+1}\zeta^{-1}.\nonumber
\end{eqnarray}

\subsection{Relation to topological open string amplitude}
The form of the vortex partition function has the close relation to
the topological open string amplitude. As the first example we consider
the vortex partition function for $U(1)$ gauge theory with
Chern-Simons level $-1/2$ with a single chiral multiplet. As already
explained at the previous subsection, the vortex partition function is given
by
\begin{eqnarray}
& &\sum_{n=0}^{\infty}w^n
x^{-\frac{n(n+1)}{2}}\prod_{k=1}^n (2 {\rm sinh} \gamma k)^{-1} \nonumber \\
&=&\sum_{n=0}^{\infty}\frac{w^n}{(1-e^{-2\gamma})(1-e^{-4\gamma})
\cdots (1-e^{-2n\gamma}) }
\end{eqnarray}
with $x=e^{-\gamma}$. Now  coinsider  the topological open
string for a Lagrangian brane in $C^3$ as explained in \cite{Gukov10, Iqbal12}\footnote{We use the refined
vertex formalism to write down the topological string partition function. For the notation, please refer to \cite{Iqbal07}.
$s_{\mu}$ in the formula denotes the Schur function.}:
\begin{eqnarray}
&& Z_{brane}(z, t, q)=\sum s_{\mu^t}(z) C_{00\mu} (t, q) \nonumber
\\ &=& \sum_{n=0}^{\infty} \frac{t^{\frac{n}{2}}z^n}{(1-q)(1-q^2)
\cdots
(1-q^n)}=\prod_{i=0}^{\infty}\frac{1}{1-q^it^{\frac{1}{2}}z}.
\end{eqnarray}

This coincides with the vortex partition function if we identify
$z=w, t=1, q=e^{-2\gamma}$. To compare with the index of the
free chiral field with the canonical $R$-charge we need the shift
$z\rightarrow z\sqrt{q}$. Then
\begin{equation}
|Z_{brane}|^2=\frac{Z_{brane}(z,q)}{Z_{brane}(\bar{z},q)}
=\frac{\prod_{n=1}^{\infty}(1-zq^{n-\frac{1}{2}})}{\prod_{n=1}^{\infty}(1-\bar{z}q^{n-\frac{1}{2}})}
\end{equation}
which coincides with the free chiral index as explained in
\cite{Iqbal12}. To compare with the free chiral index of arbitrary $R$-charge or its mirror dual, one simply
change the open string modulus $z\rightarrow z q^{\alpha}$ for a suitable $\alpha$.
Note that it's crucial to have Chern-Simons term to
match the vortex partition function with the topological open string
amplitude.

For this simple example, we generalize the matching between the homological
vortex partition function of two dimensions and the topological open string partition function
to the full 3d K-theoretic vortex partition function.\footnote{This relation
is parallel to 4d Nekrasov partition function and its 5d version.} In \cite{Gukov10}, many
more examples of the matching between the 2d vortex partition function and the topological
open string were found. We expect that this surely lifts to the matching between the 3d vortex partition
function and the topological open string. Furthermore in the homological version, 2d vortex theory is
realized as the surface operator of 4d gauge theories. We expect that this lifts to the 3d defect operator
in 5d superconformal field theories. We will work out  a simple example in the next subsection.

As a next example, we can consider $U(1)$ gauge theory with one
fundamantal and one antifundamental chiral theory. As will be  shown in
the subsection \ref{simpleseiberg},  the superconformal index of the theory
is given by \be
    I^{N_c=N_f=1} = \prod_{l=0}^\infty\frac{1-\tau^{-2}x^{2l+2}}{1-\tau^2 x^{2l}}\times
    Z_{vortex}^{N_c=N_f=1}\times Z_{anti}^{N_c=N_f=1} \ .
\ee Here $N_c$ denotes the rank of the gauge group while
$N_f=\tilde{N}_f$ denotes the number of fundamental and
antifundamental multiplets.  The vortex partition function is given by
 \be
    Z_{vortex}^{N_c=N_f=1}={\rm exp}\left[\sum_{n=1}^\infty
    \frac{1}{n}w^n\frac{(\tau^{-n}-\tau^n)x^n}{1-x^{2n}}\right] \ .\ee
    Note that the vortex part as well as perturbative part  is given
    by the free chiral index. Hence this $U(1)$ theory can again be
    written in terms of topological open string amplitude.

If one considers the more general $U(1)$ non-chiral theory with
$N_f=\tilde N_f=N$, the index can be written as
\begin{equation}
I^{N_f=\tilde N_f=N}(x,t_a,\tilde
t_a,\tau,w)=\sum_{b=1}^N\left(Z_{1-loop}^{b}Z_{vortex}^{b}\right)\times\left(Z_{1-loop,anti}^{b}Z_{anti}^{b}\right)
\end{equation}
where
\begin{eqnarray}
Z_{1-loop}^{b}&=&\prod_{k=1}^\infty\frac{\prod_{a=1(\neq b)}^{N}1-t_{b} t_a^{-1}x^{2k}}
{\prod_{a=1}^{N}1-t_{b} \tilde t_a \tau^2 x^{2(k-1)}},\\
Z_{1-loop,anti}^{b}&=&\prod_{k=1}^\infty\frac{\prod_{a=1}^{N}1-t_{b}^{-1} \tilde t_a^{-1}
\tau^{-2} x^{2k}}{\prod_{a=1(\neq b)}^{N}1-t_{b}^{-1} t_a x^{2(k-1)}},\\
Z_{vortex}^{b}&=&\sum_{n=0}^\infty\left[w\tau^{-N}x^{N}\right]^{n}
\prod_{k=1}^{n}\frac{1-t_{b} \tilde t_b \tau^2 x^{2(k-1)}}{1-x^{2k}}
\prod_{a=1(\neq b)}^{N}\frac{1-t_{b} \tilde t_a \tau^2 x^{2(k-1)}}{1-t_{b} t_a^{-1}x^{2k}},\label{vortex}\qquad
\\
Z_{anti}^{b}&=&\sum_{{\bar n}=0}^\infty\left[w^{-1}\tau^{-N}x^{N}\right]^{\bar n}\prod_{k=1}^{\bar
n}\frac{1-t_{b} \tilde t_b \tau^2
x^{2(k-1)}}{1-x^{2k}}\prod_{a=1}^{N}\frac{1-t_{b} \tilde t_a \tau^2
x^{2(k-1)}}{1-t_{b} t_a^{-1} x^{2k}}.
\end{eqnarray}
The vortex partition function \eqref{vortex} is the same as that of
\cite{Pasquetti:2011fj}. In fact, with identifications
\begin{eqnarray}
t_b t_a^{-1}&=&e^{-2\pi b D_{ab}},\\
t_b\tilde t_a\tau^2&=&e^{-2\pi b C_{ab}},\nonumber\\
x^2&=&q,\nonumber\\
w\tau^{-N}x^N&=&z,\nonumber
\end{eqnarray}
one can see that
\begin{equation}
Z_{vortex}^b=Z_{V}^{(b)}, \qquad Z_{1-loop}^b=Z_{1-loop}^{(b)}
\end{equation}
where $Z^{(b)}$'s are partition functions for the non-chiral theory
given in \cite{Pasquetti:2011fj}.

\begin{figure}
\centering
\includegraphics[scale=.7]{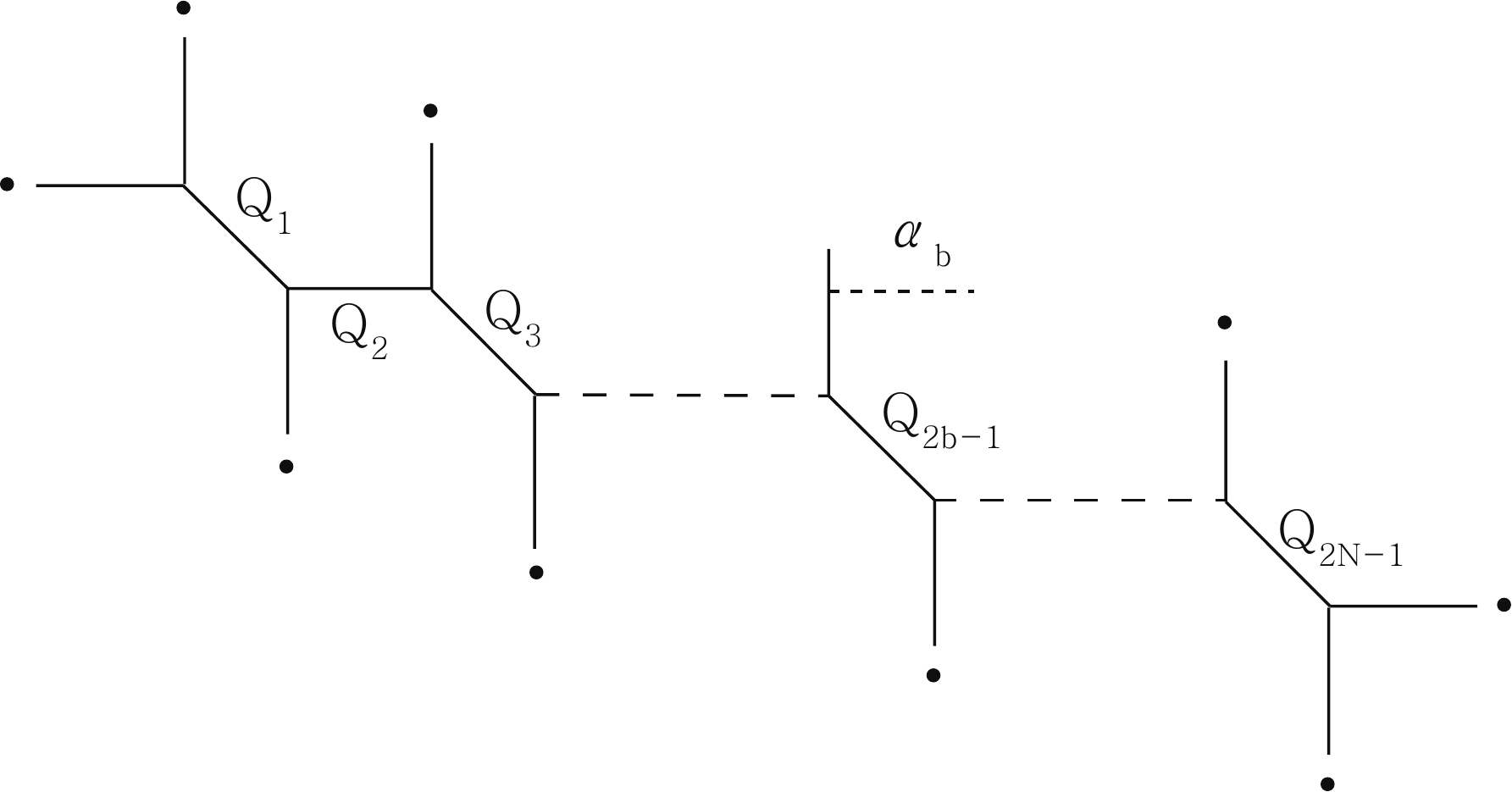}
\caption{A strip geometry.}\label{strip}
\end{figure}
As examined in \cite{Pasquetti:2011fj} one can also check that the vortex partition function $Z_{vortex}^b$ is exactly the same as the open topological string partition function on the Lagrangian brane placed at the $b$-th gauge leg of the toric diagram in Figure \ref{strip}, i.e., $\alpha_b \in \{\mathbf1^n|n=0,1\cdots\}$. The corresponding topological partition function is given by \cite{Pasquetti:2011fj,Iqbal:2004ne}
\begin{equation}
Z^b_\textrm{top}=\sum_n A_n^{(b)} z^n,
\end{equation}\begin{equation}
A^{(b)}_n\equiv\frac{\mathcal{K}^{\bullet\cdots1^n\cdots \bullet}_{\bullet\cdots \bullet} }{
\mathcal{K}^{\bullet \cdots \bullet}_{\bullet\cdots \bullet}}= \frac{1}{\prod_{k=1}^{n}(1-q^k)} \frac{\prod_{a\geq b} \prod_{k=1}^{n} (1-Q_{\alpha_b\beta_a} q^{k-1})     \prod_{a< b} \prod_{k=1}^{n} (1-Q_{\beta_a \alpha_b} q^{-(k-1)})       }{
  \prod_{a>b}\prod_{k=1}^{n} (1-Q_{\alpha_b \alpha_a} q^{k-1})  \prod_{a< b}\prod_{k=1}^{n} (1-Q_{\alpha_a \alpha_b} q^{-(k-1)}) }\, ,\\
\end{equation}
where the K\"ahler parameters are defined by:
\begin{eqnarray}
Q_{\alpha_a\alpha_{a'}}&=&\prod_{k=a}^{{a'}-1} M_k F_k\, , \nonumber\\
Q_{\alpha_a\beta_{a'}}&=&Q_{\alpha_a\alpha_{a'}} M_{a'} \, ,\\
Q_{\beta_a\alpha_{a'}}&=&Q_{\alpha_a\alpha_{a'}} M_a^{-1}\, , \nonumber
\end{eqnarray}
with $a<a'$. For a fixed $b$, if we identify the parameters as follows:
\begin{align}
z \prod_{a < b} M_a^{-1} &= w \tau^{-N} x^N, \\
Q_{\alpha_b \beta_a} &= t_b \tilde t_a \tau^2, \qquad a \geq b\\
{Q_{\beta_a \alpha_b}}^{-1} &= t_b \tilde t_a \tau^2, \qquad a < b \nonumber\\
Q_{\alpha_b \alpha_a} &= t_b t_a^{-1} x^2, \qquad a > b \nonumber\\
{Q_{\alpha_a \alpha_b}}^{-1} &= t_b t_a^{-1} x^2, \qquad a < b \nonumber
\end{align}
one immediately sees that the topological partition function $Z_\textrm{top}^b$ is the same as the vortex partition function for abelian theories that we obtained, $Z_{vortex}^b$.

\subsection{The partition function for $U(1)^N$ quiver theories and closed topological string}
More interestingly the closed string geometry on the strip geometry
considered in \cite{Pasquetti:2011fj}, for which $\alpha_b$ is now the trivial representation, has the close relation to the
5d partition function on $S^1 \times S^4$. In this case the 5d gauge theory
defined on the strip geometry is $U(1)^N$ quiver theory.

\begin{figure}
\centering
\includegraphics[scale=.7]{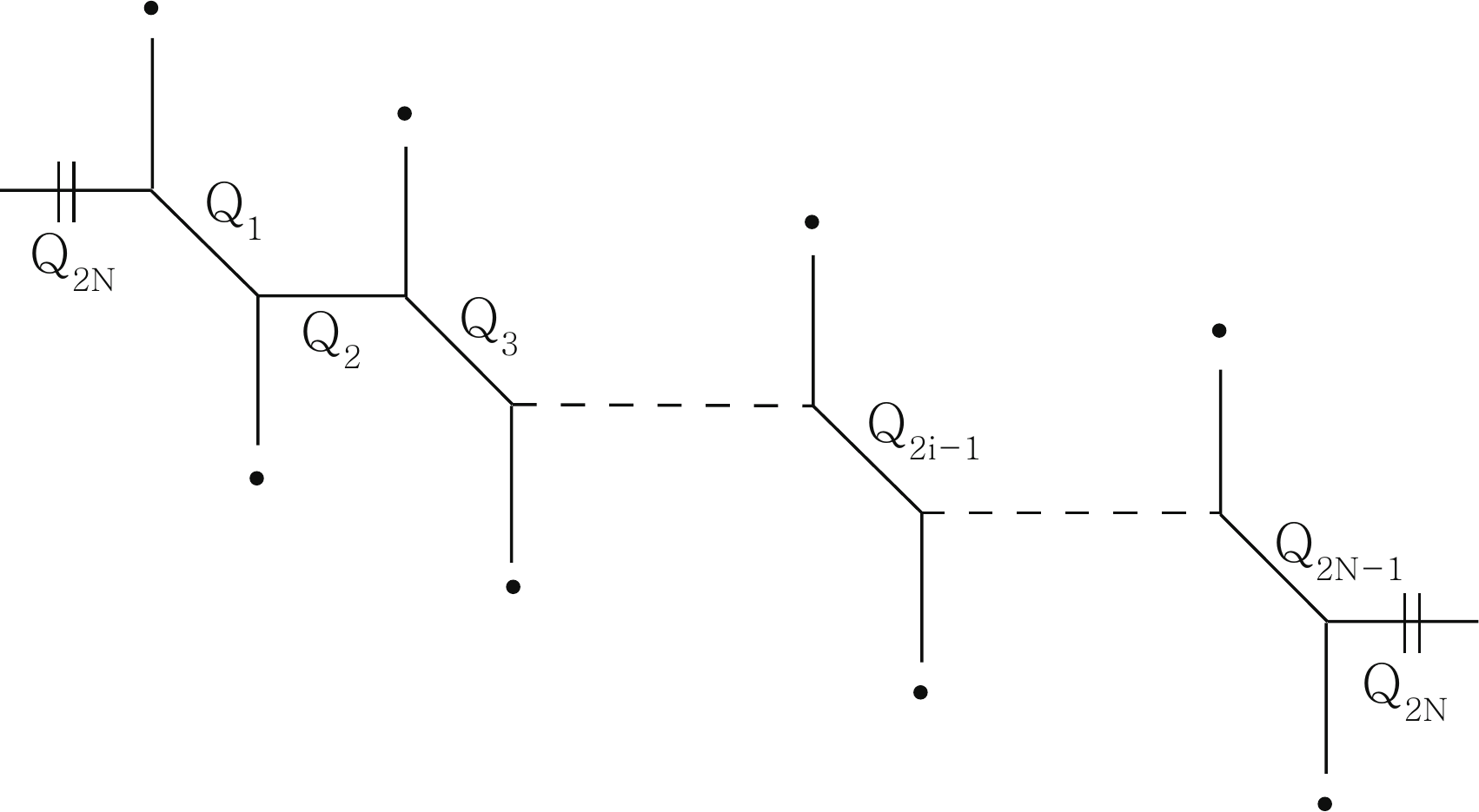}
\caption{A necklace $U(1)^N$ quiver theory.}\label{necklace}
\end{figure}
If we consider first the closed string amplitude on the strip geometry
it can be worked out using the refined topological vertex. In \cite{Awata:2009yc} a similar geometry given in Figure. \ref{necklace} was examined where the leftmost leg and the rightmost leg are identified. If we disconnect that leg we again obtain the strip geometry we are interested in. The closed string amplitude for Figure. \ref {necklace} is given by
\begin{eqnarray}
\hspace{-.5cm}&&Z_L^{inst}\left(\tilde Q;q,t\right)\\
\hspace{-.5cm}\!&\!=\!&\!\sum _{\left\{\lambda _{2\alpha }\right\}}\! \prod _{\alpha =1}^N\!
\frac{\tilde Q_{2\alpha }^{\left|\lambda _{2\alpha }\right|}\prod
_{s\in\lambda _{2\alpha -2}} \!\left(\!1\!-\!\tilde Q_{2\alpha
-1}q^{\ell_{2\alpha -2}(s)}t^{a_{2\alpha }(s)+1}\right)\!\prod
_{s\in\lambda _{2\alpha }} \!\left(1\!-\!\tilde Q_{2\alpha
-1}q^{-\ell_{2\alpha }(s)-1}t^{-a_{2\alpha -2}(s)}\right)}{\prod
_{s\in\lambda _{2\alpha }} \left(1-q^{\ell_{2\alpha
}(s)}t^{a_{2\alpha }(s)+1}\right)\left(1-q^{-\ell_{2\alpha
}(s)-1}t^{-a_{2\alpha }(s)}\right)}\nonumber
\end{eqnarray}
where $\tilde Q_\alpha=Q_\alpha \left(\frac{q}{t}\right)^{(-1)^{\alpha+1}\frac{1}{2}}$.
\footnote{If we consider the 2d partition $\lambda=\{\lambda_1 \geq \lambda_2\cdots\}$, this can be represented by a Young diagram. We draw the $\lambda_1$ boxes on the leftmost column and $\lambda_2$ boxes on the next-leftmost column and so on.
For  an element $s=(i,j) \in \lambda$,  $a(s)$ denotes the boxes on the right and $l(s)$ denotes the boxes on top, i.e.,
$a(i,j)=\lambda_j^t-i, l(i,j)=\lambda_i-j$. For more details, refer to \cite{Iqbal07}.} As in the other examples of geometric engineering, this closed
topological string amplitude leads to the Nekrasov instanton
partition function. In 5-dimensions full instanton partition
function was not worked out for theories with bifundamental fields.
However Nekrasov partition function of such quiver in four-dimension
was worked out  in \cite{Fucito:2004gi}. One can see that this
can be obtained from the closed string amplitude.  The
unrefined version of the amplitude is obtained by setting $t=q$:
\begin{eqnarray}
&&Z_L^{inst}\left(\tilde Q_\alpha;q,q\right)\nonumber\\
&=&\sum _{\left\{\lambda _{2\alpha }\right\}} \prod _{\alpha =1}^N \tilde Q_{2\alpha }^{\left|\lambda _{2\alpha }\right|}
\prod _{s\in\lambda _{2\alpha}} \frac{\left(1-\tilde Q_{2\alpha +1}q^{\ell_{2\alpha}(s)+a_{2\alpha +2}(s)+1}\right)
\left(1-\tilde Q_{2\alpha -1}q^{-\ell_{2\alpha }(s)-a_{2\alpha -2}(s)-1}\right)}{\left(1-q^{\ell_{2\alpha }(s)
+a_{2\alpha }(s)+1}\right)\left(1-q^{-\ell_{2\alpha }(s)-a_{2\alpha }(s)-1}\right)}\nonumber\\
&=&\sum _{\left\{\lambda _{2\alpha }\right\}} \prod _{\alpha =1}^N \left(\tilde Q_{2\alpha }
\tilde Q_{2\alpha+1}^{1/2}\tilde Q_{2\alpha-1}^{1/2}\right)^{\left|\lambda _{2\alpha }\right|}
q^{\sum_{s\in\lambda_{2\alpha}}[a_{2\alpha +2}(s)-a_{2\alpha -2}(s)]/2}\nonumber\\
&&\times\prod _{s\in\lambda _{2\alpha}}
\frac{\sinh\frac{\beta}{2}[\hbar
h_{2\alpha,2\alpha+2}(s)+M_{2\alpha+1}] \sinh\frac{\beta}{2}[\hbar
h_{2\alpha,2\alpha-2}(s)-M_{2\alpha-1}]}{\sinh^2\frac{\beta}{2}\hbar
h_{2\alpha,2\alpha}(s)}
\end{eqnarray}
where we defined that $\tilde Q_\alpha\equiv e^{-\beta
M_\alpha}$, $q\equiv e^{-\beta\hbar}$. $\beta M_\alpha$ and
$\beta\hbar$ here are the five-dimensional chemical potentials while
$M_\alpha$ and $\hbar$ are the four-dimensional parameters.
$h_{\alpha,\beta}(s)$ is the hook length defined by
$h_{\alpha,\beta}(s)=\ell_\alpha(s)+a_\beta(s)+1$. In order to
obtain the four-dimensional partition function, one would take
$\beta\rightarrow0$:
\begin{eqnarray}
&&\left.Z_L^{inst}\left(\tilde Q_\alpha;q,q\right)\right|_{\beta\rightarrow0}\nonumber\\
&=&\sum _{\left\{\lambda _{2\alpha }\right\}} \prod _{\alpha
=1}^N\prod _{s\in\lambda _{2\alpha}} \frac{[\hbar
h_{2\alpha,2\alpha+2}(s)+M_{2\alpha+1}] [\hbar
h_{2\alpha,2\alpha-2}(s)-M_{2\alpha-1}]}{[\hbar
h_{2\alpha,2\alpha}(s)]^2},
\end{eqnarray}
which is the same as the partition function for quiver theories
given in \cite{Fucito:2004gi} with identifications
$M_{2\alpha+1}=a_{2\alpha}-a_{2\alpha+2}+m$ where $m$ denotes the mass of the
bifundamentals.\footnote{The hook
length $h_{\alpha,\beta}(s)$ is denoted by $\ell_{\alpha\beta}(s)$
in \cite{Fucito:2004gi}.} Thus it is quite reasonable that the above topolgical string
amplitude gives the 5d Nekrasov partition function for $U(1)^N$
quiver theories. One can cut the leftmost leg and the the rightmost leg, which are identified, by taking $Q_{2N}\rightarrow0$. This give rise to the closed string amplitude for the strip geometry we original considered. The amplitude is given by
\begin{eqnarray}
&&\left.Z_L^{inst}\left(\tilde Q;q,t\right)\right|_{Q_{2N}\rightarrow0}\nonumber\\
&=&\sum _{\left\{\lambda _{2\alpha }\right\}}\prod_{\alpha=1}^{N-1}\tilde Q_{2\alpha }^{\left|\lambda _{2\alpha }\right|}\prod _{s\in\lambda _{2}} \left(1-\tilde Q_{1}q^{-\ell_{2}(s)-1}t^{-a_{\emptyset}(s)}\right)\prod _{s\in\lambda _{2N-2}} \left(1-\tilde Q_{2N-1}q^{\ell_{2N-2}(s)}t^{a_{\emptyset}(s)+1}\right)\nonumber\\
&&\times \frac{\prod _{\alpha =2}^{N-1}\prod _{s\in\lambda _{2\alpha -2}} \left(1-\tilde Q_{2\alpha -1}q^{\ell_{2\alpha -2}(s)}t^{a_{2\alpha }(s)+1}\right)\prod _{s\in\lambda _{2\alpha }} \left(1-\tilde Q_{2\alpha -1}q^{-\ell_{2\alpha }(s)-1}t^{-a_{2\alpha -2}(s)}\right)}{\prod _{\alpha =1}^{N-1}\prod _{s\in\lambda _{2\alpha }} \left(1-q^{\ell_{2\alpha }(s)}t^{a_{2\alpha }(s)+1}\right)\left(1-q^{-\ell_{2\alpha }(s)-1}t^{-a_{2\alpha }(s)}\right)}\nonumber\\
\end{eqnarray}
where $a_\emptyset(s=(i,j))=-i$.

One might wonder since abelian theory is trivial in
5d so that its nonperturbative part is also trivial. However abelian
theories can have small instantons and it's quite subtle how to
include them. For example if we consider 5d $U(1)~ \mathcal N=2^*$ theory and if
we define its nonperturbative completion to give the Nekrasov
partition function, 5d partition function of $U(1)~ \mathcal N=2^*$ theory on $S^5$ gives  the index of single M5 brane in 6d \cite{6dindex}. 

The general structure of the 5d index worked out at \cite{KKL12} has the
structure
\begin{equation}
\int da PE(f_{mat}(x,y,e^{ia},t)+f_{vec}(x,y,e^{ia})) |I_{inst}(x,y,e^{ia},t,q)|^2
\end{equation}
where $da$ is the Haar measure for the gauge group, $PE$ denotes Plesthystic exponetial, which gives the one-loop determinant and
$I_{inst}$ is Nekrasov instanton partition function.  $x,y$ is the chemical potential for Cartans of Lorentz symmetry
$SU(2)_1 \times SU(2)_2\subset SO(5), x=e^{-\gamma_1}, y=e^{-\gamma_2}$ and $t$ is the usual chemical potential
for the flavor symmetry. Here $SU(2)_1$ is also twisted with $SU(2)_R$ R-symmetry.
Finally $q$ is introduced to track the instanton number. Thus for $U(1)^N$ quiver 5d partition
function has the same form where $I_{inst}$ is now identified with closed
string amplitude. This is consistent with the recent proposal by
\cite{Iqbal12}.

In addition, the perturbative part is also factorized and the whole index can be written as
\begin{equation}
I=\int da |I_{pert}(x,y,e^{ia}, t)I_{inst}(x,y,e^{ia},t,q)|^2.
\end{equation}
Now let's check if perturbative part matches. In the refined vertex formalism,
the preturbative part is automatically built in. In our case, it is given by \cite{Awata:2009yc}
\begin{equation}
Z_{pert}={\rm exp} \left(-\sum_{n=1}^{\infty}\frac{1}{n}\frac{\sum_{\alpha}\tilde{Q}_{2\alpha-1}^n}
{(t^{\frac{n}{2}}-t^{-\frac{n}{2}})(q^{\frac{n}{2}}-q^{-\frac{n}{2}})}\right).
\end{equation}
This should match the one-loop determinant of the bifundamental fields.
The general expression for the one-loop determinant for the matter fields are given by
\begin{equation}
f_{mat}(x,y,a)=\frac{x}{(1-xy)(1-x/y)}\sum_w (e^{-i\vec{w}\cdot\vec{\a}}+e^{i\vec{w}\cdot\vec{\a}})
\end{equation}
where $w$ is the weight of the representation. We suppres the chemical potential of the flavor symmetry. We can see that it has the explicit factorized structure
and we can just look for $e^{-i\vec{w}\cdot\vec{\a}}$ part to compare with the topological string expression.
For bifundamentals, we have
\begin{equation}
f_{pert}=\frac{x}{(1-xy)(1-x/y)}(e^{-i(\a_1-\a_2)}+e^{-i(\a_2-\a_3)}+\cdots +e^{-i(\a_N-\a_1)}).
\end{equation}
This coincides with the corresponding topological string expression if we identify
\begin{equation}
x=\sqrt{tq},\qquad y=\sqrt{\frac{q}{t}},\qquad\tilde{Q}_{2k-1}=e^{-i(\a_k-\a_{k-1})}.
\end{equation}

Furthermore since the open string amplitude was obtained by
introducing the Lagrangian brane in the strip geometry, and this
leads to the 3d index of the nontrivial SCFT, it is natural to
expect that introducing Lagrangian brane corresponds to introducing
the surface operator in 5d $U(1)^N$ quiver theory. This is the T-dual of
the Hanany-Witten set up for the surface operator in 4-dimension
so we expect this is the 3d defect of the 5d theory.

This lead to an interesting lesson that apparently trivial 5d theory\footnote{One way to see the 5d index computation
of the this theory is to regard it as a twisted partition function on $S^1\times S^4$.} can have
nontrivial defect operator, which corresponds to nontrivial 3d
SCFT. Furthermore we saw that the partition function of 5d theory
with the defect operator matches the closed+open string amplitude
since the the vertex partition function appearing at \cite{Pasquetti:2011fj},
is normalized by the closed string partition function.  The vortex partition function
has the structure
\begin{equation}
Z^b_{vortex}=\sum_n z^n \frac{K(1^n)}{K(0)}
\end{equation}
where $K(1^n)$ is the string partition function with the insertion of the brane
with the representation $(1^n)$ while $K(0)$ denotes the string partition function
with the trivial representation, i.e., the closed string partition function.

\section{$\mathcal{N}=2$ Seiberg-like dualities}
\subsection{Simple cases}\label{simpleseiberg}
In this section we consider Seiberg-like (or Aharony duality) for three dimensional
$U(N)$ gauge theories with $\mathcal{N}=2$ supersymmetries proposed
in \cite{Aharony:1997gp}. The duality relates two gauge theories
which we call the ``original" theory and the ``dual" theory. Two
dual theories have different gauge groups and matter contents but
they flow to the same theory in the infrared.

The original theory is a $U(N)$ gauge theory which consists of $N_f$
fundamental chiral multiplets $Q_a$ and $N_f$ anti-fundamental
chiral multiplets $\tilde{Q}^a$ as well as $U(N)$ vector multiplets.
This theory has no superpotential. On the other hand, the dual
theory is a $U(N_f-N)$ gauge theory with $N_f$ pairs of fundamental
$q^a$ and anti-fundamental $\tilde{q}_a$ chiral multiplets. In
addition, the dual theory contains gauge singlet chiral multiplets,
$M_a{}^b$ and $V_\pm$, and they couple to the charged matters
through the superpotential, $W = q^a M_a{}^b \tilde{q}_b +
V_+\tilde{V}_- + V_- \tilde{V}_+$. Here $\tilde{V}_\pm$ are chiral
superfields corresponding to monopole operators which parametrize
the Coulomb branch of the dual theory. The global symmetry of both
theories is $SU(N_f)\times SU(N_f)$ flavor symmetry acting on the
fundamental and anti-fundamental matters, respectively, times
$U(1)_A\times U(1)_T$, where $U(1)_A$ is an axial symmetry rotating
fundamental and anti-fundamental matters by the same phase and $U(1)_T$ is a
topological symmetry whose current is given by $*{\rm Tr} F$.

Under the duality, mesonic operators $Q_a\tilde{Q}^b$ and monopole
operators with topological charges $\pm1$ of the original theory are
mapped to singlet fields $M_a{}^b$ and $V_\pm$ of the dual theory,
respectively. This duality map together with the superpotential $W$
determines global charge assignment of chiral fields of the dual
theory.

The superconformal indices for several dual pairs have been computed.
The indices are expanded by conformal dimensions of
BPS operators and show agreement between BPS spectra of two dual
theories at some leading orders. Here we present factorized
representations of superconformal indices for simple cases that
shows 3d Seiberg-like dualities in a clearer way.

Let us first consider the $U(1)$ gauge theory with $N_f=1$ flavor
which would give the simplest duality model. The proposed dual
theory is the $U(0)$ theory, i.e. non-gauge theory, with chiral
multiplets $M$ and $V_\pm$. After vortex-antivortex factorization,
the superconformal index of the original theory is given by \be
    I^{N=N_f=1} = \prod_{l=0}^\infty\frac{1-\tau^{-2}x^{2l+2}}{1-\tau^2 x^{2l}}
    \times Z_{vortex}^{N=N_f=1}\times Z_{ anti}^{N=N_f=1} \ .
\ee The vortex index is the sum over all vortex number $n$'s.
After some calculation it can be written as a Plethystic exponential
form
\be
    Z_{{vortex}}^{N=N_f=1}=\sum_{n=0}^\infty(-w)^n\prod_{k=1}^n\frac{\tau^{-1}x^{-(k-1)}-\tau x^{k-1}}{x^{-(k-1-n)}-x^{k-1-n}}={\rm exp}\left[\sum_{n=1}^\infty\frac{1}{n}w^n\frac{(\tau^{-n}-\tau^n)x^n}{1-x^{2n}}\right] \ .
\ee
We checked the last identity up to the order $\mathcal{O}(w^9)$. The
anti-vortex index is easily obtained from the vortex index by
replacing $w$ to $w^{-1}$. Moreover it turns out that the
superconformal index of $N=N_f=1$ theory can be rewritten as a
simple Plethystic exponential form
\be\label{index-(1,1)}
    I^{N=N_f=1} \!&\!=\!&\! {\rm exp}\left[\sum_{n=1}^\infty \frac{1}{n}f(x^n,\tau^n,w^n)\right] \ , \\
    f(x,\tau,w) \!&\!=\!&\! \frac{\tau^2x^{2\Delta_Q}-\tau^{-2}x^{2-2\Delta_Q}}{1-x^2}+\frac{\tau^{-1}x^{1-\Delta_Q}-\tau x^{1+\Delta_Q}}{1-x^2}(w+w^{-1}) \ ,\nn
\ee where we restored $R$-charge $\Delta_Q$ of the chiral boson $Q$ of
the original theory. Amazingly this form of the index is exactly the
same as the superconformal index of the dual theory. The function
$f$ is identical to the single letter index in the dual theory. As
the chiral field $M$ of the dual theory is identified with the meson
operator $Q\tilde{Q}$ of the original theory, its $R$-charge and
$U(1)_A$ charge are $2\Delta_Q$ and $+2$ respectively, and therefore
the letter index of $M$ is given by the first term of $f$. The
second term of $f$ comes from the letter contribution of dual chiral
multiplets $V_\pm$ which is mapped to monopole operators with
$U(1)_T$ charges $\pm1$. In general, zero point energies and
$U(1)_A$ charges of monopole operators with GNO charge
$(\pm1,0,\cdots,0)$ for $N=N_f$ theories are \be
    \epsilon_0 &=& N_f(1-\Delta_Q)-(N-1) = 1-N\Delta_Q \ , \\
    b_{U(1)_A} &=& -N_f = -N\nn
\ee from \eqref{components}. One can then see that the single letter index of
$V_\pm$ for $N=N_f=1$ case agrees with the second term of $f$.

This theory is known to be mirror-dual to the XYZ theory \cite{AhaHan97}.
The chiral fields $M,V_\pm$ in the dual theory correspond to $X,Y,Z$ fields of the superpotential $W=-MV_+V_-$,
so that they should have $R$-charges $\Delta_M=\Delta_V=\frac{2}{3}$.
As shown in \cite{Jafferis10}, the $R$-charge of the original chiral field is determined to be $\Delta_Q=\frac{1}{3}$ in IR,
and therefore one can see from \eqref{index-(1,1)} that the dual chiral fields have the correct $R$-charges in the IR fixed point.

More generally, one can express the superconformal indice for $U(N)$
gauge theories with $N_f=N$ fundamental and anti-fundamental matters
in duality manifest forms using the factorization. The dual theory
is a $U(0)$ theory with chiral multiplet $M_a^{\ b}$ and $V_\pm$.
The vortex indices for $N=N_f$ theories reduce to Plethystic
exponential forms \be
    Z_{{vortex}}^{N=N_f}\!&\!=\!&\! \sum_{\vec{n}=0}^\infty(-w)^n\prod_{i,j}^N\prod_{k=1}^{n_i}
    \frac{t_i^{-1/2}\tilde{t}_j^{-1/2}\tau^{-1}x^{-(k-1)}-t_i^{1/2}\tilde{t}_j^{1/2}\tau x^{k-1}}{t_i^{-1/2}t_j^{1/2}x^{-(k-1-n_i)}-t_i^{1/2}t_j^{-1/2}x^{k-1-n_i}} \nn \\
    \!&\!=\!&\!{\rm exp}\left[\sum_{n=1}^\infty\frac{1}{n}w^n\frac{(\tau^{-Nn}-\tau^{Nn})x^n}{1-x^{2n}}\right] \ .
\ee We explicitly checked the last identity for some low values of
$n$ and $N$. Together with the antivortex partition function, this can be interpreted as the multi-particle index for singlet chiral fields $V_\pm$ of the dual theory. All of the $t$ dependence are cancelled out, which is
expected since $V_\pm$ are the flavor singlets. Restoring $R$-charge
by shifting $\tau\rightarrow \tau x^{\Delta_Q}$, one can check the
chiral field $V_+$ has correct $R$-charge, $1-N\Delta_Q$, and $U(1)_A$
charge, $-N$. Then the superconformal index after combining the
perturbative part can also be rewritten as duality manifest form
\be\label{index-(N,N)}
    I^{N=N_f} \!&\!=\!&\! \prod_{i,j}^{N}\prod_{l=0}^\infty\frac{1-t_i^{-1}\tilde{t}_j^{-1}\tau^{-2}x^{2l+2-2\Delta_Q}}{1-t_i\tilde{t}_j\tau^2 x^{2l+2\Delta_Q}} \times Z_{vortex}^{N=N_f} \times Z_{anti}^{N=N_f} \nn \\
    \!&\!=\!&\!  {\rm exp}\left[\sum_{n=1}^\infty \frac{1}{n}f_{N=N_f}(x^n,t^n,\tilde{t}^n,\tau^n,w^n)\right] \,,  \\
    f^{N=N_f} \!&\!=\!&\! \sum_{i,j}^{N}\frac{t_i\tilde{t}_j\tau^2x^{2\Delta_Q}-t_i^{-1}\tilde{t}_j^{-1}\tau^{-2}x^{2-2\Delta_Q}}{1-x^2}
    +\frac{\tau^{-N}x^{1-N\Delta_Q}-\tau^{N}x^{1+N\Delta_Q}}{1-x^2}(w+w^{-1})\nn
\ee This precisely agrees with superconformal index of the dual
theory with $N\!\times\! N$ chiral fields $M_i^{\ j}$ and two chiral
fields $V_\pm$.
When $N>1$, the dual theories flow to free theories.
One can check it first for $N=2$, where the $Z$-extrimization of \cite{Jafferis10} determines $R$-charge of the original chiral fields as $\Delta_Q=\frac{1}{4}$.
Then $R$-charges of the dual chiral fields are fixed to be $\Delta_M=\Delta_V = \frac{1}{2}$ and so the dual theory is obviously free.
For $N>2$, it seems to be impossible to have free dual theory by adjusting the original $R$-charge $\Delta_Q$.
However, as we see from the index formula \eqref{index-(N,N)}, the index of the IR conformal theory is written as that of non-interacting free fields
and therefore IR degrees of freedom can carry new $U(1)$ charges for accidental symmetry which emerges only at the IR fixed point.
The UV $R$-symmetry then mixes with this extra $U(1)$ symmetry so that the dual chiral fields $M,V_\pm$ become free fields in IR.

Now we consider further generalization to $N_f>N$ theories. Unlike
the previous cases which are mostly free theories, dual theories are
now interacting gauge theories. Let us first consider $U(1)$ gauge
theory with $N_f=2$ pairs of fundamental and anti-fundamental
matters. The dual theory is also $U(1)$ gauge theory with $N_f=2$
flavors, but has additional $2\!\times\!2$ chiral fields $M_a{}^b$
and two chiral fields $V_\pm$. The superconformal index of the
original theory is \be\label{index-for-N=1-N_f=2}
    I^{(1,2)} \!&\!=\!&\! \sum_{\sigma(t)}Z_{pert}^{(1,2)}(x,\sigma(t),\tilde{t},\tau) \times Z_{vortex}^{(1,2)}(x,\sigma(t),\tilde{t},\tau,w)
    \times Z_{anti}^{(1,2)}(x,\sigma(t),\tilde{t},\tau,w^{-1}) \,, \nn \\
    Z_{pert}^{(1,2)}\!&\!=\!&\! \prod_{l=0}^\infty\left[\frac{1-t_1t_2^{-1}x^{2l+2}}{1-t_1^{-1}t_2x^{2l}}
    \prod_{a=1}^2\frac{1-t_1^{-1}\tilde{t}_a^{-1}\tau^{-2}x^{2l+2-2\Delta_Q}}{1-t_1\tilde{t}_a\tau^2 x^{2l+2\Delta_Q}}\right] \,, \nn \\
    Z_{ vortex}^{(1,2)}\!&\!=\!&\! \sum_{n=0}^\infty (-w)^n \prod_{k=1}^n \frac{\prod_{a=1}^{2}(t_1^{-1/2}\tilde{t}_a^{-1/2}\tau^{-1} x^{-(k-1)-\Delta_Q}-t_1^{1/2}\tilde{t}_a^{1/2}\tau x^{k-1+\Delta_Q})
    }{(x^{-(k-1-n)}-x^{k-1-n})(t_1^{-1/2}t_2^{1/2}x^{-k}-t_1^{1/2}t_2^{-1/2}x^k)}
\ee and $Z_{\rm anti}^{(1,2)}=Z_{vortex}^{(1,2)}(w\rightarrow w^{-1})$ where $I^{(N,N_f)}$ denotes
the index of the original theory with $U(N)$ gauge group and $N_f$
flavors, and $\sigma(t)$ runs over permutations of $\{t_1,t_2\}$.
In fact this index also has the duality manifest expression. The
perturbative part $Z_{pert}^{(1,2)}$ with exchange of $t_a$'s
can be rewritten as \be\label{pert-index-(1,2)}
    \hspace{-.5cm}Z_{pert}^{(1,2)}(t_1\leftrightarrow t_2)\!&\!=\!&\!\prod_{l=0}^\infty\!\!\left[\!\frac{1\!-\!t_1^{-1}t_2x^{2l+2}}{1\!-\!t_1t_2^{-1}x^{2l}}
    \prod_{a=1}^2\!\frac{1\!-\!t_1\tilde{t}_a\tau^{2}x^{2l+2\Delta_Q}}{1\!-\!t_1^{-1}\tilde{t}_a^{-1}\tau^{-2} x^{2l+2(1-\Delta_Q)}}
    \cdot\prod_{a,b}^2\!\frac{1\!-\!t_a^{-1}\tilde{t}_b^{-1}\tau^{-2} x^{2l+2-2\Delta_Q}}{1-t_a\tilde{t}_b\tau^2 x^{2l+2\Delta_Q}}\!\right] \nn \\
    \hspace{-.5cm}\!&\!=\!&\! \tilde{Z}_{ pert}^{(1,2)} \times \prod_{l=0}^\infty\prod_{a,b}^2\frac{1\!-\!t_a^{-1}\tilde{t}_b^{-1}\tau^{-2} x^{2l+2-2\Delta_Q}}{1-t_a\tilde{t}_b\tau^2 x^{2l+2\Delta_Q}}
\ee where $\tilde{Z}^{(1,2)}_{pert} \equiv Z^{(1,2)}_{
pert}\left(t\!\rightarrow \!t^{-1},\tilde{t}\!\rightarrow
\!\tilde{t}^{-1},\tau\!\rightarrow\! \tau^{-1},\Delta_Q\!
\rightarrow \!1\!-\!\Delta_Q\right)$. We shall identify
$\tilde{Z}^{(1,2)}_{pert}$ to the perturbative part of charged
chiral fields $q^a,\tilde{q}_a$ in the dual theory. Also the second
infinity product term in the second line of
\eqref{pert-index-(1,2)} will be identified with the index
contribution of the meson field $M_a{}^b$ of the dual theory.
Similarly, we define the dual vortex index as
$\tilde{Z}^{(1,2)}_{vortex} \equiv Z^{(1,2)}_{
vortex}\left(t\!\rightarrow \!t^{-1},\tilde{t}\!\rightarrow
\!\tilde{t}^{-1},\tau\!\rightarrow\! \tau^{-1},\Delta_Q\!
\rightarrow \!1\!-\!\Delta_Q\right)$ and find that \be
    Z_{vortex}^{(1,2)}(t_1,t_2) = \tilde{Z}^{(1,2)}_{vortex}(t_2,t_1)\times{\rm exp}\!\!\left[\sum_{n=1}^\infty\frac{1}{n}w^n\frac{\tau^{-2n}x^{2n(1-\Delta_Q)}\!-\!\tau^{2n}x^{2n\Delta_Q}}{1-x^{2n}}\right]
\ee We checked this identity by expanding both sides with vortex
number up to $\mathcal{O}(w^6)$. The Plethystic exponential term on
the right hand side corresponds to the index contribution from
chiral fields $V_\pm$ of the dual theory. Finally, collecting all
the result, the original index becomes \be
    I^{(1,2)} \!&\!=\!&\! \tilde{I}^{(1,2)}\times  {\rm exp}\left[\sum_{n=1}^\infty \frac{1}{n}f^{(1,2)}(x^n,t^n,\tilde{t}^n,\tau^n,w^n)\right]
\ee where \be
    \tilde{I}^{(1,2)} \!&\!=\!&\!  \sum_{\sigma(t)}\tilde{Z}_{pert}^{(1,2)}(\sigma(t)) \times \tilde{Z}_{vortex}^{(1,2)}(\sigma(t))
    \times \tilde{Z}_{anti}^{(1,2)}(\sigma(t)) \,,  \\
    f^{(1,2)}  \!&\!=\!&\! \sum_{a,b}^2\frac{t_a\tilde{t}_b\tau^2x^{2\Delta_Q}-t_a^{-1}\tilde{t}_b^{-1}\tau^{-2}x^{2-2\Delta_Q}}{1-x^2}
    +\frac{\tau^{-2}x^{2(1-\Delta_Q)}-\tau^{2}x^{2\Delta_Q}}{1-x^2}(w+w^{-1})\nn
\ee This is exactly the same as the superconformal index of the dual
theory, which is a $U(1)$ gauge theory with charged chiral
multiplets $q^a,\tilde{q}_a$, singlet chiral multiplets $M_a{}^b$
and $V_\pm$. The superpotential $W$ implies that $R$-charges for $q,
\tilde{q}$ are $1\!-\!\Delta_Q$ and other charges are opposite to
$Q, \tilde{Q}$ of the original theory. Thus the index
$\tilde{I}^{(1,2)}$ encodes the contributions from the chiral
multiplets $q, \tilde{q}$. One can also check that the single letter
index $f^{(1,2)}$ represents the letter indices for $M_a{}^b$ and
$V_\pm$ with correct $R$-charge and global charges.

\subsection{General cases}
One can generalize the $N=1,N_f=2$ example in the previous subsection to general $N,N_f$ in the same way. The index contribution of the singlet matters $M_a{}^b,V_\pm$ is straightforward, and the contribution of $q,\tilde q$ is obtained by replacing $N\rightarrow N-N_f$, $t,\tilde t\rightarrow t^{-1},\tilde t^{-1}$, $\tau\rightarrow \tau^{-1}x$ in the original index. Thus, the superconformal index for the dual theory is given by
\begin{eqnarray}
\hspace{-.4cm}&&I(x=e^{-\gamma},t=e^{iM},\tilde t=e^{i\tilde M},\tau=e^{i\mu},w)\nonumber\\
\hspace{-.4cm}\!&\!=\!&\!\left[\!\prod_{k=0}^\infty\!\!\left(\!\prod_{a,b=1}^{N_f}\!\!\frac{1\!-\!t_a^{-1}\tilde t_b^{-1}\tau^{-2} x^{2+2k}}{1\!-\!t_a\tilde t_b\tau^2 x^{2k}}\!\right)\!\!\!\left(\!\frac{1\!-\!w^{-1}\tau^{N_f} x^{1-N_f+N+2k}}{1\!-\!w\tau^{-N_f} x^{N_f-N+1+2k}}\!\right)\!\!\left(\!\frac{1\!-\!w\tau^{N_f} x^{1-N_f+N+2k}}{1\!-\!w^{-1}\tau^{-N_f} x^{N_f-N+1+2k}}\!\right)\!\right]\nonumber\\
\hspace{-.4cm}&&\times\sum_{1\leq b_1<\cdots<b_{N_f-N}\leq N_f}\nonumber\\
\hspace{-.4cm}&&\textstyle\left\{\!\!\left(\!\prod_{i,j=1(i\neq j)}^{N_f-N}\!-2\sinh\frac{iM_{b_i}\!-\!iM_{b_j}}{2}\!\right)\!\!\left[\!\prod_{j=1}^{N_f\!-\!N}\!\prod_{k=0}^\infty\!\!\left(\!\frac{\prod_{a=1(\neq b_j)}^{N_f}1-t_{b_j}^{-1} t_a x^{2+2k}}{\prod_{a=1}^{N_f}1-t_{b_j}^{-1} \tilde t_a^{-1} \tau^{-2} x^{2+2k}}\!\right)\!\!\left(\!\frac{\prod_{a=1}^{N_f}1-t_{b_j} \tilde t_a \tau^{2} x^{2k}}{\prod_{a=1(\neq b_j)}^{N_f}1-t_{b_j} t_a^{-1} x^{2k}}\!\right)\!\right]\right.\nonumber\\
\hspace{-.4cm}&&\times\!\sum_{\vec n=0}^\infty\!\!\textstyle\left[\!(-w)^{\sum n_j}\!\!\!\left(\!\!\prod_{j=1}^{N_f-N}\!\prod_{k=1}^{n_j}\!\!\frac{\prod_{a=1}^{N_f}2\sinh\frac{i\tilde M_a+iM_{b_j}+2i\mu+2\gamma k}{2}}{\left(\!\prod_{i=1}^{N_f\!-\!N}\!2\sinh\!\frac{-iM_{b_i}\!+\!iM_{b_j}\!+\!2\gamma(k\!-\!1\!-\!n_i)}{2}\right)\!\!\left(\!\prod_{a\in\{b_j\}^c}\!2\sinh\frac{-iM_a\!+\!iM_{b_j}\!+\!2\gamma k}{2}\right)}\!\!\right)\!\!\right]\nonumber\\
\hspace{-.4cm}&&\times\!\left.\sum_{\vec{\bar{n}}=0}^\infty\!\!\textstyle\left[\!(-w)^{-\sum\bar n_j}\!\!\!\left(\!\!\prod_{j=1}^{N_f-N}\!\prod_{k=1}^{\bar n_j}\!\!\frac{\prod_{a=1}^{N_f}\!2\sinh\!\frac{i\tilde M_a\!+\!iM_{b_j}\!+\!2i\mu\!+\!2\gamma k}{2}}{\left(\!\prod_{i=1}^{N_f\!-\!N}\!2\sinh\!\frac{-iM_{b_i}\!+\!iM_{b_j}\!+\!2\gamma(k\!-\!1\!-\!\bar n_i)}{2}\!\right)\!\!\left(\!\prod_{a\in\{b_j\}^c}\!2\sinh\!\frac{-iM_a\!+\!iM_{b_j}\!+\!2\gamma k}{2}\!\right)}\right)\!\!\right]\!\!\right\}.\nonumber\\
\hspace{-.4cm}
\end{eqnarray}
where $\{b_j\}^c=\{1,\cdots,N_f\}-\{b_1,\cdots,b_{N_f-N}\}$. The second line comes from the singlet matters $M_a{}^{b}$ and $V_\pm$. The fourth line is the perturbative part coming from $q^a$ and $\tilde q_a$, which will be called $\tilde Z_{pert}$. The last two lines are vortex and antivortex parts, which will be called $\tilde Z_{vortex}$ and $\tilde Z_{anti}$.

With a little algebra one can show that the following expression holds:
\begin{eqnarray}
&&\prod_{b\in\{b_j\}}\prod_{k=0}^\infty\left(\frac{\prod_{a=1(\neq b)}^{N_f}1-t_{b} t_a^{-1}x^{2+2k}}{\prod_{a=1}^{N_f}1-t_{b} \tilde t_a \tau^2 x^{2k}}\right)\left(\frac{\prod_{a=1}^{N_f}1-t_{b}^{-1} \tilde t_a^{-1} \tau^{-2} x^{2+2k}}{\prod_{a=1(\neq b)}^{N_f}1-t_{b}^{-1} t_a x^{2k}}\right)\nonumber\\
\!&\!=\!&\!\left(\prod_{a,b=1}^{N_f}\prod_{k=0}^\infty\frac{1-t_a^{-1}\tilde t_b^{-1}\tau^{-2} x^{2+2k}}{1-t_a\tilde t_b\tau^2 x^{2k}}\right)\nonumber\\
&&\times\!\left(\!\frac{\prod_{a,b\in\{b_j\}^c(a\neq b)}1\!-\!t_{b} t_a^{-1}}{\prod_{a,b\in\{b_j\}(a\neq b)}1\!-\!t_{b}^{-1} t_a}\right)\!\!\!\left[\!\prod_{b\in\{b_j\}^c}\prod_{k=0}^\infty\!\!\left(\!\frac{\prod_{a=1(\neq b)}^{N_f}1\!-\!t_{b}^{-1} t_a x^{2+2k}}{\prod_{a=1}^{N_f}1\!-\!t_{b}^{-1} \tilde t_a^{-1} \tau^{-2} x^{2+2k}}\!\right)\!\!\left(\!\frac{\prod_{a=1}^{N_f}1\!-\!t_{b} \tilde t_a \tau^2 x^{2k}}{\prod_{a=1(\neq b)}^{N_f}1\!-\!t_{b} t_a^{-1} x^{2k}}\!\right)\!\!\right]\nonumber\\
\end{eqnarray}
for an arbitrary subset $\{b_j\}\subset\{1,\cdots,N_f\}$. It teaches us that we can write the perturbative part of the original index as follows:
\begin{eqnarray}
Z_{pert}^{\{b_j\}}\left(x,t,\tilde t,\tau\right)&=&\tilde Z_{pert}^{\{b_j\}^c}\left(x,t,\tilde t,\tau\right)\times\left(\prod_{a,b=1}^{N_f}\prod_{k=0}^\infty\frac{1-t_a^{-1}\tilde t_b^{-1}\tau^{-2} x^{2+2k}}{1-t_a\tilde t_b\tau^2 x^{2k}}\right)\nonumber\\
&=&\tilde Z_{pert}^{\{b_j\}^c}\left(x,t,\tilde t,\tau\right)\times\exp\left[\sum_{n=1}^\infty\frac{1}{n}f_M\left(x^n,t^n,\tilde t^n,\tau^n\right)\right],\\
f_M(x,t,\tilde t,\tau)&=&\sum_{a,b=1}^{N_f}\frac{t_a\tilde t_b\tau^2-t_a^{-1}\tilde t_b^{-1}\tau^{-2}x^2}{1-x^2}.
\end{eqnarray}
One would note that $f_M$ is exactly the letter index for the Mesons $M_a{}^b$. Now every term of the remaining part has a nonzero power of $w$ except 1. In addition, $Z_{vortex}$ only has positive powers of $w$ while $Z_{anti}$ has negative powers of $w$. Therefore, we conjecture that the following identities hold:
\begin{eqnarray}
Z_{vortex}^{\{b_j\}}(x,t,\tilde t,\tau,w)\!&\!=\!&\!\tilde Z_{vortex}^{\{b_j\}^c}\left(x,t,\tilde t,\tau,w\right)\!\times\!\left(\prod_{k = 0}^\infty \frac{1-w\tau^{N_f} x^{1-N_f+N+2k}}{1-w\tau^{-N_f} x^{N_f-N+1+2k}}\right)\nonumber\\
\!&\!=\!&\!\tilde Z_{vortex}^{\{b_j\}^c}\left(x,t,\tilde t,\tau,w\right)\!\times\!\exp\!\left[\sum_{n=1}^\infty\frac{1}{n}f_{+}\left(x^n,t^n,\tilde t^n,\tau^n,w^n\right)\right],\qquad
\end{eqnarray}
\begin{eqnarray}
Z_{anti}^{\{b_j\}}(x,t,\tilde t,\tau,w)&=&\tilde Z_{anti}^{\{b_j\}^c}\left(x,t,\tilde t,\tau,w\right)\times\left(\prod_{k = 0}^\infty \frac{1-w^{-1}\tau^{N_f} x^{1-N_f+N+2k}}{1-w^{-1}\tau^{-N_f} x^{N_f-N+1+2k}}\right)\nonumber\\
&=&\tilde Z_{anti}^{\{b_j\}^c}\left(x,t,\tilde t,\tau,w\right)\!\times\!\exp\!\left[\sum_{n=1}^\infty\frac{1}{n}f_{-}\left(x^n,t^n,\tilde t^n,\tau^n,w^n\right)\right],\qquad
\end{eqnarray}
\begin{eqnarray}
f_{\pm}(x,t,\tilde t,\tau,w)&=&w^\pm\frac{\tau^{-N_f}x^{N_f-N+1}-\tau^{N_f}x^{1-N_f+N}}{1-x^2},
\end{eqnarray}
which is the generalization of the numerically tested identities for special $N,N_f$ above.
 We also check validity of these formulae by extensive numerical computation. Note that $f_{+}+f_{-}=f_{V_+}+f_{V_-}$ where $f_{V_\pm}$ are the letter indices for the singlets $V_\pm$, which have nonzero topological charges:
\begin{equation}
f_{V_\pm}(x,t,\tilde t,\tau,w)=\frac{w^\pm\tau^{-N_f}x^{N_f-N+1}-w^\mp\tau^{N_f}x^{1-N_f+N}}{1-x^2}
\end{equation}
In fact, $f_+$ gets the contribution from the scalar of $V_+$ and the fermion of $V_-$ while $f_-$ gets the contribution from the fermion of $V_+$ and the scalar of $V_-$. For both the perturbative part and the vortex part, the contribution with certain choice of $\{b_j\}_\textrm{orig}$ for the original theory is exactly the same as the contribution with the complementary choice of $\{b_j\}_\textrm{dual}={\{b_j\}_\textrm{orig}}^c$. Summing over all possible choices of $\{b_j\}$, the total indices for both theories are thus the same for any $N$ and $N_f\geq N$. Note that the perturbative contribution of $Q_a$ and $\tilde Q^a$ maps to that of $q^a,\tilde q_a$ and the contribution of $M_a{}^b$ while the vortex and antivortex contributions of $Q_a,\tilde Q^a$ map to those of $q^a,\tilde q_a$ and the contribution of $V_\pm$.

\subsection{$\mathcal{N}=4$ Seiberg-like duality and mirror symmetry}
$\mathcal{N}=4$ Seiberg-like dualities were proposed in \cite{Kapustin10-2, KKKL12} based on brane configuration of Type IIB string theory.
Under the duality, a $U(N)$ gauge theory with $N_f$ fundamental hypermultiplets is conjectured to be dual to another $U(N_f-N)$ theory with $N_f$ hypers in the infrared.

In this section we consider the simplest example of $\mathcal{N}=4$ Seiberg-like duality.
At low energy, an $\mathcal{N}=4$ $U(1)$ gauge theory with a fundamental hypermultiplet and a free theory of one hypermultiplet flow to the same theory. This is also the simplest example of the mirror symmetry. The free hypermultiplet is so called the twisted hypermultiplet in the context of mirror symmetry.
As two theories are simple enough, we can easily compare two superconformal indices of them and check this duality conjecture.
In the case at hand, the $U(1)$ gauge multiplet of the original theory couples to one fundamental and anti-fundamental chiral matters
while the adjoint chiral matter is decoupled from it.
So there is a similarity between this $U(1)$ theory and $\mathcal{N}=2$ $U(1)$ gauge theory with $N_f=\tilde{N}_f=1$ chiral matters up to the decoupled adjoint chiral.
In fact, once we assign the correct global charges to $\mathcal{N}=2$ fields, it is easy to write the superconformal index of $\mathcal{N}=4$ $U(1)$ gauge theory
using $\mathcal{N}=2$ result.
Then the superconformal index for $U(1)$ theory with $N_f=1$ fundamental hyper after factorization becomes
\be
    I^{\mathcal{N}=4}_{N=N_f=1} \!&\!=\!&\! I^{\mathcal{N}=2}_{N=N_f=1}(\Delta_Q\!=\!\tfrac{1}{2},\tau=y^{1/2})\times
    {\rm exp}\!\!\left[\sum_{n=1}^\infty\frac{1}{n}\frac{x^n(y^{-n} - y^{n})}{1-x^{2n}}\right] \nn \\
    \!&\!=\!&\! {\rm exp}\left[\sum_{n=1}^\infty\frac{1}{n}\frac{y^{-n/2}x^{n/2} - y^{n/2}x^{3n/2}}
    {1-x^{2n}}(w^n+w^{-n})\right] \label{mirror}
\ee Here $I^{\mathcal{N}=2}_{N=N_f=1}$ is the index of
(\ref{index-(1,1)}) for $\mathcal{N}=2$ theory, and we set $R$-charge
of bosonic fields to be $\Delta_Q\!=\!\frac{1}{2}$ and introduced
the chemical potential $y$ for the off-diagonal $U(1)_A$ of
$SU(2)_L\times SU(2)_R=SO(4)$ R-symmetry. The exponential term on
the right hand side of the first line is from the adjoint chiral
multiplet. The final expression is written as the Plethystic
exponential of one free hypermultiplet that agrees with the duality
proposal. In the dual theory $w$ is interpreted as the $U(1)$ flavor
chemical potential. Note that this also perfectly matches with
mirror symmetry. Under the mirror symmetry the monopole operator of
the $U(1)$ is mapped to the twisted free hypermultiplet. Note that
$w$ at eq. (\ref{mirror}) is the vortex number, which is nothing but
the monopole charge. In the mirror side this is mapped to the charge
of the flavor symmetry of the free hypermultiplet. The detailed
exploration of the mirror symmetry and $\mathcal{N}=4$ Seiberg-like duality in terms of the factorization
will appear elsewhere.

\section{Concluding remarks}
There would be manifold generalizations one can pursue related to the current work.
The first one is the direct proof of the factorization using the localization. For the 2d partition function,
it is explicitly worked out in \cite{benini12}. Certainly it is more desirable to more general gauge groups and
general matters, which will have applications for Seiberg-like dualities for classical groups with two index
matters. This was explored in \cite{Kapustin11}.

For simple cases, we already saw the vortex partition function coincides with the corresponding topological
open string amplitude. Such pattern will hold for more general cases and it would be desirable to work out
explicitly. In \cite{Gukov10}, the 2d vortex arises as the surface operator of the 4d supersymmetric gauge theory
and we expect that this will be lifted to the 3d defect to the 5d SCFT. The 3d SCFT realized as the IR limit of the
3d gauge theory flows to the 2d CFT upon the dimensional reduction. Thus many of the properties of 2d CFTs such as
conformal blocks and $tt*$ equations will be lifted
to the corresponding 3d CFTs, which is interesting to explore.
In the same spirit, the relation between 3d mirror symmetry and 2d mirror symmetry would be worked out in similar way.
2d mirror symmetry in the nonabelian gauge group setup is explored recently\cite{Morrison12, Gomis12} and it would be interesting to find its
relation to 3d mirror symmetry.

Finally the proof of the duality such as Seiberg-like duality, mirror symmetry
will be greatly simplified with the factorized form of the index and it is worth attempting analytic proof
of the index equality for dual pairs.

%Seiberg dualities proposed in \cite{Aharony:1997gp} relate two different 3d gauge theories with $\mathcal{N}=2$ supersymmetry in the infrared limit.
%Here we focus on the duality for $U(N)$ gauge theories.
%The duality relates the ``original" theory with $U(N)$ gauge symmetry to its ``dual" theory with $U(N_f\!-\!N)$ gauge symmetry.
%The original theory consists of $N_f$ fundamental chiral fields $Q_i$ and $N_f$ anti-fundamental chiral fields $\tilde{Q}^{\tilde{j}}$ as well as the $U(N)$ vector multiplet.
%Its global symmetry is $SU(N_f)\times \tilde{SU(N_f)}\times U(1)_A\times U(1)_T$ where the first
%The theory has Higgs branch of moduli space parametrized by the gauge invariant meson field $M_i^{\ \tilde{j}}=Q_i\tilde{Q}^{\tilde{j}}$
%and Coulomb branch of moduli space parametrized by monopole chiral fields $V_\pm$ carrying $U(1)_T$ topological charge $\pm 1$.

\vskip 0.5cm  \hspace*{-0.8cm} {\bf\large Acknowledgements} \vskip
0.2cm

\hspace*{-0.75cm}  We thank Dongmin Gang, Kimyeong Lee, Sungjay Lee and Seok Kim for useful discussions.
 J.P. was supported by the National Research
Foundation of Korea (NRF) grant funded by the Korea government (MEST) with the
Grants No. 2009-0085995, 2012-009117, 2012-046278 (JP) and 2005-0049409 (JP) through
the Center for Quantum Spacetime (CQUeST) of Sogang University. JP is also supported
by the POSTECH Basic Science Research Institute Grant and appreciates APCTP for its
stimulating environment for research.

\appendix
\section{Factorization: nonabelian cases}
In this section we are going to derive the factorized form of the superconformal index for a $U(N)$ gauge theory with $N_f$ fundamental and $\tilde N_f$ antifundamental flavors. At first the superconformal index is given by
\begin{equation}
I(x,t,w,\kappa)=\sum_{\vec m\in\mathbb Z^{N}/S_{N}}\oint\prod_j\frac{dz_j}{2\pi i z_j}\frac{1}{|\mathcal W_m|}w^{\sum_j m_j}e^{-S_{CS}(a,m)}Z_{gauge}(x,z,m)\prod_\Phi Z_\Phi(x,t,z,m)
\end{equation}
where
\begin{equation}
e^{-S_{CS}(a,m)}=e^{-i\textrm{Tr}_{CS}(a+\pi) m},
\end{equation}
\begin{equation}
Z_{gauge}\left(x,z=e^{ia},m\right)=\prod_{\alpha\in ad(G)}x^{-|\alpha(m)|/2}\left(1-e^{i\alpha(a)}x^{|\alpha(m)|}\right),
\end{equation}
\begin{eqnarray}
&&Z_\Phi\left(x,t,z=e^{ia},m\right)\nonumber\\
&=&\prod_{\rho\in R_\Phi}\left(x^{(1-\Delta_\Phi)}e^{-i\rho(a+\pi)}\prod_a t_a^{-f_a(\Phi)}\right)^{|\rho(m)|/2}\frac{\left(e^{-i\rho(a)}\prod t_a^{-f_a(\Phi)}x^{|\rho(m)|+2-\Delta_\Phi};x^2\right)_\infty}{\left(e^{i\rho(a)}\prod t_a^{f_a(\Phi)}x^{|\rho(m)|+\Delta_\Phi};x^2\right)_\infty}.\nonumber\\
\end{eqnarray}
$(a;q)_n$ is the $q$-Pochhamers symbol defined by
\begin{equation}
(a;q)_n=\prod_{k=0}^{n-1}\left(1-a q^k\right),\qquad|q|<1.
\end{equation}
We have included the nontrivial phase shift $a_j\rightarrow a_j+\pi$ for nonzero magnetic flux vacuua. If one considers a $U(N)$ gauge theory with $N_f$ fundamental and $\tilde N_f$ antifundamental chiral multiplets,
\begin{equation}
e^{-S_{CS}(a,m)}=\prod_{j=1}^{N}(-z_j)^{-\kappa m_j},
\end{equation}
\begin{equation}
Z_{gauge}(x,z,m)=\prod_{\substack{i,j=1\\(i\neq j)}}^{N}x^{-|m_i-m_j|/2}\left(1-z_iz_j^{-1}x^{|m_i-m_j|}\right),
\end{equation}
\begin{eqnarray}
&&\prod_\Phi Z_\Phi\left(x,t,\tilde t,\tau,z,m\right)\nonumber\\
&=&x^{(1-\Delta_\Phi)(N_f+\tilde N_f)\sum|m_j|/2}\left[\prod_{j=1}^{N}(-z_j)^{-(N_f-\tilde N_f)|m_j|/2}\right]\tau^{-(N_f+\tilde N_f)\sum|m_j|/2}\nonumber\\
&&\prod_{j=1}^{N}\prod_{k=0}^\infty\left(\prod_{a=1}^{N_f}\frac{1-z_j^{-1}t_a^{-1}\tau^{-1}x^{|m_j|+2-\Delta_\Phi+2k}}{1-z_j t_a\tau x^{|m_j|+\Delta_\Phi+2k}}\right)\left(\prod_{a=1}^{\tilde N_f}\frac{1-z_j \tilde t_a^{-1}\tau^{-1}x^{|m_j|+2-\Delta_\Phi+2k}}{1-z_j^{-1}\tilde t_a\tau x^{|m_j|+\Delta_\Phi+2k}}\right)\nonumber\\
\end{eqnarray}
where $t_a$ and $\tilde t_a$ correspond to fugacities for $SU(N_f)\times SU(\tilde N_f)$; $\tau$ is a fugacity for $U(1)_A$. One expects that $\kappa+(N_f+\tilde N_f)/2$ should be an integer due to the quantization of the effective CS level. In addition, we will set $\Delta_\Phi=0$, which can be restored by  deforming $\tau\rightarrow \tau x^{\Delta_\Phi}$. The infinite product only makes sense for $|x|<1$. Thus, if we assume $|t_a\tau|,|\tilde t_a\tau|<1$, which can be extended by analytic continuation after integration, poles from the antifundamental part lie inside the integration contour. Indeed, the integrand also has a pole at the origin, which makes the integration difficult, for $N_f\geq\tilde N_f$. Fortunately for $N_f>\tilde N_f$ one could change the integration variables $z_j\rightarrow 1/z_j$ to exclude the pole at the origin and would take poles from the fundamental part, which are now inside the contour, instead of the poles from the antifundamental part. For $N_f=\tilde N_f$, however, one should take account of the pole at the origin.

Here we are dealing with the $N_f>\tilde N_f$ case first. Changing the variables $z_j\rightarrow 1/z_j$ is equivalent to summing residues at poles outside the contour, which come from the fundamental part: $z_j=t_{b_j}^{-1}\tau^{-1}x^{-|m_j|-2l_j}$ for $b_j=1,\cdots,N_f$ and $l_j=0,1,\cdots$. After performing the contour integration the index is given by
\begin{eqnarray}
&&I^{N_f>\tilde N_f}(x,t,\tilde t,\tau,w,\kappa)\nonumber\\
&=&\sum_{\vec m\in\mathbb Z^{N}/S_{N}}\sum_{b_1,\cdots,b_{N}=1}^{N_f}\sum_{\vec l=0}^\infty\frac{1}{|\mathcal W_m|}(-1)^{-\kappa\sum m_j-(N_f-\tilde N_f)|m_j|/2}w^{\sum m_j}\left(\prod_{j=1}^{N}t_{b_j}^{\kappa m_j+(N_f-\tilde N_f)|m_j|/2}\right)\nonumber\\
&&\tau^{\kappa\sum m_j-\tilde N_f\sum|m_j|}x^{\kappa\sum m_j(|m_j|+2l_j)-\sum_{i\neq j}|m_i-m_j|/2+(N_f+\tilde N_f)\sum|m_j|/2+(N_f-\tilde N_f)\sum\left(|m_j|^2+2|m_j|l_j\right)/2}\nonumber\\
&&\left(\prod_{\substack{i,j=1\\(i\neq j)}}^{N}1-t_{b_i}^{-1}t_{b_j}x^{|m_i-m_j|-|m_i|+|m_j|-2l_i+2l_j}\right)\nonumber\\
&&\left[\prod_{j=1}^{N_C}\left(\frac{\prod_{a=1}^{N_f}\prod_{k=0}^\infty1-t_{b_j} t_a^{-1}x^{2|m_j|+2l_j+2+2k}}{\prod_{\substack{a=1,k=0\\((a,k)\neq(b_j,l_j))}}^{N_f,\infty}1-t_{b_j}^{-1} t_a x^{-2l_j+2k}}\right)\left(\prod_{a=1}^{\tilde N_f}\prod_{k=0}^\infty\frac{1-t_{b_j}^{-1} \tilde t_a^{-1} \tau^{-2} x^{-2l_j+2+2k}}{1-t_{b_j}\tilde t_a \tau^2 x^{2|m_j|+2l_j+2k}}\right)\right].\nonumber\\
\end{eqnarray}
Let us look at $\prod_{\substack{i,j=1\\(i\neq j)}}^{N}1-t_{b_i}^{-1}t_{b_j}x^{|m_i-m_j|-|m_i|+|m_j|-2l_i+2l_j}$, $\prod_{\substack{a=1,k=0\\((a,k)\neq(b_j,l_j))}}^{N_f,\infty}1-t_{b_j}^{-1} t_a x^{-2l_j+2k}$ and $\prod_{a=1}^{\tilde N_f}\prod_{k=0}^\infty1-t_{b_j}^{-1} \tilde t_a^{-1} \tau^{-2} x^{-2l_j+2+2k}$. They can be rewritten as follows:
\begin{eqnarray}
\hspace{-1cm}&&\prod_{\substack{i,j=1\\(i\neq j)}}^{N}1-t_{b_i}^{-1}t_{b_j}x^{|m_i-m_j|-|m_i|+|m_j|-2l_i+2l_j}\nonumber\\
\hspace{-.6cm}&=&\prod_{i<j}^{N}\left(1-t_{b_i}^{-1}t_{b_j}x^{m_i-m_j-|m_i|+|m_j|-2l_i+2l_j}\right)\left(1-t_{b_i}t_{b_j}^{-1}x^{m_i-m_j+|m_i|-|m_j|+2l_i-2l_j}\right)\nonumber\\
\hspace{-1cm}&=&\prod_{i<j}^{N}\left(-x^{m_i-m_j}\right)\left(t_{b_i}^{1/2}t_{b_j}^{-1/2}x^{(l_i+|m_i|/2-m_i/2)-(l_j+|m_j|/2-m_j/2)}-t_{b_i}^{-1/2}t_{b_j}^{1/2}x^{-(l_i+|m_i|/2-m_i/2)+(l_j+|m_j|/2-m_j/2)}\right)\nonumber\\
\hspace{-1cm}&&\quad\times\left(t_{b_i}^{1/2}t_{b_j}^{-1/2}x^{(l_i+|m_i|/2+m_i/2)-(l_j+|m_j|/2+m_j/2)}-t_{b_i}^{-1/2}t_{b_j}^{1/2}x^{-(l_i+|m_i|/2+m_i/2)+(l_j+|m_j|/2+m_j/2)}\right),
\end{eqnarray}
\begin{eqnarray}
&&\prod_{\substack{a=1,k=0\\((a,k)\neq(b_j,l_j))}}^{N_f,\infty}1-t_{b_j}^{-1} t_a x^{-2l_j+2k}\nonumber\\
&=&\left(\prod_{a=1}^{N_f}\prod_{k=0}^{l_j-1}1-t_{b_j}^{-1} t_a x^{-2l_j+2k}\right)\left(\prod_{\substack{a=1,k=0\\((a,k)\neq(b_j,0))}}^{N_f,\infty}1-t_{b_j}^{-1} t_a x^{2k}\right)\nonumber\\
&=&\left(\prod_{a=1}^{N_f}\prod_{k=0}^{l_j-1}-t_{b_j}^{-1} t_a x^{-2l_j+2k}\right)\left(\prod_{a=1}^{N_f}\prod_{k=0}^{l_j-1}1-t_{b_j} t_a^{-1} x^{2+2k}\right)\left(\prod_{\substack{a=1,k=0\\((a,k)\neq(b_j,0))}}^{N_f,\infty}1-t_{b_j}^{-1} t_a x^{2k}\right),\nonumber\\
\end{eqnarray}
\begin{eqnarray}
&&\prod_{a=1}^{\tilde N_f}\prod_{k=0}^\infty1-t_{b_j}^{-1} \tilde t_a^{-1} \tau^{-2} x^{-2l_j+2+2k}\nonumber\\
&=&\left(\prod_{a=1}^{\tilde N_f}\prod_{k=0}^{l_j-1}1-t_{b_j}^{-1} \tilde t_a^{-1} \tau^{-2} x^{-2l_j+2+2k}\right)\left(\prod_{a=1}^{\tilde N_f}\prod_{k=0}^\infty1-t_{b_j}^{-1} \tilde t_a^{-1} \tau^{-2} x^{2+2k}\right)\nonumber\\
&=&\left(\prod_{a=1}^{\tilde N_f}\prod_{k=0}^{l_j-1}-t_{b_j}^{-1} \tilde t_a^{-1} \tau^{-2} x^{-2l_j+2+2k}\right)\left(\prod_{a=1}^{\tilde N_f}\prod_{k=0}^{l_j-1}1-t_{b_j} \tilde t_a \tau^2 x^{2k}\right)\left(\prod_{a=1}^{\tilde N_f}\prod_{k=0}^\infty1-t_{b_j}^{-1} \tilde t_a^{-1} \tau^{-2} x^{2+2k}\right).\nonumber\\
\end{eqnarray}
We have dropped the absolute value symbol of $|m_i-m_j|$ by assuming $m_1\geq\cdots\geq m_{N}$. Using these one can rewrite the index as follows:
\begin{eqnarray}
\hspace{-2cm}&&I^{N_f>\tilde N_f}(x,t,\tilde t,\tau,w,\kappa)\nonumber\\
\hspace{-2cm}&=&\sum_{\vec m\in\mathbb Z^{N}/S_{N}}\sum_{b_1,\cdots,b_{N}=1}^{N_f}\sum_{\vec l=0}^\infty\frac{1}{|\mathcal W_m|}(-1)^{-\kappa\sum m_j-(N_f-\tilde N_f)\sum|m_j|/2}w^{\sum m_j}\left(\prod_{j=1}^{N}t_{b_j}^{\kappa m_j+(N_f-\tilde N_f)|m_j|/2}\right)\nonumber\\
\hspace{-2cm}&&\tau^{\kappa\sum m_j-\tilde N_f\sum|m_j|}x^{\kappa\sum m_j(|m_j|+2l_j)-\sum_{i<j}(m_i-m_j)+(N_f+\tilde N_f)\sum|m_j|/2+(N_f-\tilde N_f)\sum\left(|m_j|^2+2|m_j|l_z\right)/2}\nonumber\\
\hspace{-2cm}&&\left[\prod_{i<j}^{N}\left(-x^{m_i-m_j}\right)\left(t_{b_i}^{1/2}t_{b_j}^{-1/2}x^{(l_i+|m_i|/2-m_i/2)-(l_j+|m_j|/2-m_j/2)}-t_{b_i}^{-1/2}t_{b_j}^{1/2}x^{-(l_i+|m_i|/2-m_i/2)+(l_j+|m_j|/2-m_j/2)}\right)\right.\nonumber\\
\hspace{-2cm}&&~~~~~\times\left.\left(t_{b_i}^{1/2}t_{b_j}^{-1/2}x^{(l_i+|m_i|/2+m_i/2)-(l_j+|m_j|/2+m_j/2)}-t_{b_i}^{-1/2}t_{b_j}^{1/2}x^{-(l_i+|m_i|/2+m_i/2)+(l_j+|m_j|/2+m_j/2)}\right)\right]\nonumber\\
\hspace{-2cm}&&\left(\prod_{j=1}^{N_C}\frac{\left(\prod_{a=1}^{N_f}\prod_{k=0}^\infty1-t_{b_j} t_a^{-1}x^{2+2k}\right)/\left(\prod_{a=1}^{N_f}\prod_{k=0}^{|m_j|+l_j-1}1-t_{b_j} t_a^{-1}x^{2+2k}\right)}{\left(\prod_{a=1}^{N_f}\prod_{k=0}^{l_j-1}-t_{b_j}^{-1} t_a x^{-2l_j+2k}\right)\left(\prod_{a=1}^{N_f}\prod_{k=0}^{l_j-1}1-t_{b_j} t_a^{-1} x^{2+2k}\right)\left(\prod_{\substack{a=1,k=0\\((a,k)\neq(b_j,0))}}^{N_f,\infty}1-t_{b_j}^{-1} t_a x^{2k}\right)}\right.\nonumber\\
\hspace{-2cm}&&~~~~~\times\left.\frac{\left(\prod_{a=1}^{\tilde N_f}\prod_{k=0}^{l_j-1}-t_{b_j}^{-1} \tilde t_a^{-1} \tau^{-2} x^{-2l_j+2+2k}\right)\left(\prod_{a=1}^{\tilde N_f}\prod_{k=0}^{l_j-1}1-t_{b_j} \tilde t_a \tau^2 x^{2k}\right)\left(\prod_{a=1}^{\tilde N_f}\prod_{k=0}^\infty1-t_{b_j}^{-1} \tilde t_a^{-1} \tau^{-2} x^{2+2k}\right)}{\left(\prod_{a=1}^{\tilde N_f}\prod_{k=0}^{\infty}1-t_{b_j} \tilde t_a \tau^2 x^{2k}\right)/\left(\prod_{a=1}^{\tilde N_f}\prod_{k=0}^{|m_j|+l_j-1}1-t_{b_j} \tilde t_a \tau^2 x^{2k}\right)}\right)\nonumber\\
\hspace{-2cm}&=&\sum_{\vec m\in\mathbb Z^{N}/S_{N}}\sum_{b_1,\cdots,b_{N}=1}^{N_f}\sum_{\vec l=0}^\infty\frac{1}{|\mathcal W_m|}(-1)^{N(N-1)/2-\kappa\sum m_j-(N_f-\tilde N_f)\sum(|m_j|+2l_j)/2}\nonumber\\
\hspace{-2cm}&&w^{\sum m_j}\left(\prod_{j=1}^{N}t_{b_j}^{\kappa m_j+(N_f-\tilde N_f)(|m_j|+2l_j)/2}\right)\tau^{\kappa\sum m_j-\tilde N_f\sum(|m_j|+2l_j)}\nonumber\\
\hspace{-2cm}&&x^{\kappa\sum m_j(|m_j|+2l_j)+(N_f+\tilde N_f)\sum(|m_j|+2l_j)/2+(N_f-\tilde N_f)\sum\left[(|m_j|+l_j)^2+l_j^2\right]/2}\nonumber\\
\hspace{-2cm}&&\left[\prod_{i<j}^{N}\left(t_{b_i}^{1/2}t_{b_j}^{-1/2}x^{(l_i+|m_i|/2-m_i/2)-(l_j+|m_j|/2-m_j/2)}-t_{b_i}^{-1/2}t_{b_j}^{1/2}x^{-(l_i+|m_i|/2-m_i/2)+(l_j+|m_j|/2-m_j/2)}\right)\right.\nonumber\\
\hspace{-2cm}&&~~~~~\times\left.\left(t_{b_i}^{1/2}t_{b_j}^{-1/2}x^{(l_i+|m_i|/2+m_i/2)-(l_j+|m_j|/2+m_j/2)}-t_{b_i}^{-1/2}t_{b_j}^{1/2}x^{-(l_i+|m_i|/2+m_i/2)+(l_j+|m_j|/2+m_j/2)}\right)\right]\nonumber\\
\hspace{-2cm}&&\left[\prod_{j=1}^{N}\left(\prod_{k=0}^\infty\frac{\prod_{a=1(\neq b_j)}^{N_f}1-t_{b_j} t_a^{-1}x^{2+2k}}{\prod_{a=1}^{\tilde N_f}1-t_{b_j} \tilde t_a \tau^2 x^{2k}}\right)\left(\prod_{k=0}^\infty\frac{\prod_{a=1}^{\tilde N_f}1-t_{b_j}^{-1} \tilde t_a^{-1} \tau^{-2} x^{2+2k}}{\prod_{a=1(\neq b_j)}^{N_f}1-t_{b_j}^{-1} t_a x^{2k}}\right)\right.\nonumber\\
\hspace{-2cm}&&~~~~~\times\left.\left(\prod_{k=0}^{|m_j|+l_j-1}\frac{\prod_{a=1}^{\tilde N_f}1-t_{b_j} \tilde t_a \tau^2 x^{2k}}{\prod_{a=1}^{N_f}1-t_{b_j} t_a^{-1}x^{2+2k}}\right)\left(\prod_{k=0}^{l_j-1}\frac{\prod_{a=1}^{\tilde N_f}1-t_{b_j} \tilde t_a \tau^2 x^{2k}}{\prod_{a=1}^{N_f}1-t_{b_j} t_a^{-1} x^{2+2k}}\right)\right].
\end{eqnarray}
In order to proceed further one should rearrange the summations. First the summation $\sum_{\vec m\in\mathbb Z^{N}/S_{N}}$ is replaced by $\sum_{\vec m\in\mathbb Z^{N}}\frac{Sym}{N!}$. Thanks to the symmetry $|m_j|+l_j\leftrightarrow l_j$ one can rearrange the summations as $\sum_{\vec m\in\mathbb Z^{N}}\sum_{\vec l=0}^\infty=\sum_{\vec n=0}^\infty\sum_{\vec{\bar n}=0}^\infty$ where $n_j\equiv l_j+\frac{|m_j|}{2}+\frac{m_j}{2}$, $\bar n_j\equiv l_j+\frac{|m_j|}{2}-\frac{m_j}{2}$ and can write the index in the factorized form:
\begin{eqnarray}
&&I^{N_f>\tilde N_f}(x,t,\tilde t,\tau,w,\kappa)\nonumber\\
&=&\frac{(-1)^{N(N-1)/2}}{N!}\sum_{b_1,\cdots,b_{N}=1}^{N_f}\sum_{\vec n=0}^\infty\sum_{\vec{\bar n}=0}^\infty\nonumber\\
&&(-1)^{-\kappa\sum (n_j-\bar n_j)-(N_f-\tilde N_f)\sum(n_j+\bar n_j)/2}w^{\sum(n_j-\bar n_j)}\left(\prod_{j=1}^{N}t_{b_j}^{\kappa(n_j-\bar n_j)+(N_f-\tilde N_f)(n_j+\bar n_j)/2}\right)\nonumber\\
&&\tau^{\kappa\sum(n_j-\bar n_j)-\tilde N_f\sum(n_j+\bar n_j)}x^{\kappa\sum(n_j^2-\bar n_j^2)+(N_f+\tilde N_f)\sum(n_j+\bar n_j)/2+(N_f-\tilde N_f)\sum\left(n_j^2+\bar n_j^2\right)/2}\nonumber\\
&&\left[\prod_{i<j}^{N}\left(t_{b_i}^{1/2}t_{b_j}^{-1/2}x^{\bar n_i-\bar n_j}-t_{b_i}^{-1/2}t_{b_j}^{1/2}x^{-(\bar n_i-\bar n_j)}\right)\left(t_{b_i}^{1/2}t_{b_j}^{-1/2}x^{n_i-n_j}-t_{b_i}^{-1/2}t_{b_j}^{1/2}x^{-(n_i-n_j)}\right)\right]\nonumber\\
&&\left[\prod_{j=1}^{N}\left(\prod_{k=0}^\infty\frac{\prod_{a=1(\neq b_j)}^{N_f}1-t_{b_j} t_a^{-1}x^{2+2k}}{\prod_{a=1}^{\tilde N_f}1-t_{b_j} \tilde t_a \tau^2 x^{2k}}\right)\left(\prod_{k=0}^\infty\frac{\prod_{a=1}^{\tilde N_f}1-t_{b_j}^{-1} \tilde t_a^{-1} \tau^{-2} x^{2+2k}}{\prod_{a=1(\neq b_j)}^{N_f}1-t_{b_j}^{-1} t_a x^{2k}}\right)\right.\nonumber\\
&&~~~~~~\times\left.\left(\prod_{k=0}^{n_j-1}\frac{\prod_{a=1}^{\tilde N_f}1-t_{b_j} \tilde t_a \tau^2 x^{2k}}{\prod_{a=1}^{N_f}1-t_{b_j} t_a^{-1}x^{2+2k}}\right)\left(\prod_{k=0}^{\bar n_j-1}\frac{\prod_{a=1}^{\tilde N_f}1-t_{b_j} \tilde t_a \tau^2 x^{2k}}{\prod_{a=1}^{N_f}1-t_{b_j} t_a^{-1} x^{2+2k}}\right)\right]\nonumber
\end{eqnarray}
\begin{eqnarray}
&=&\frac{(-1)^{N(N-1)/2}}{N!}\sum_{b_1,\cdots,b_{N}=1}^{N_f}\left\{\left[\prod_{j=1}^{N}\prod_{k=0}^\infty\left(\frac{\prod_{a=1(\neq b_j)}^{N_f}1-t_{b_j} t_a^{-1}x^{2+2k}}{\prod_{a=1}^{\tilde N_f}1-t_{b_j} \tilde t_a \tau^2 x^{2k}}\right)\left(\frac{\prod_{a=1}^{\tilde N_f}1-t_{b_j}^{-1} \tilde t_a^{-1} \tau^{-2} x^{2+2k}}{\prod_{a=1(\neq b_j)}^{N_f}1-t_{b_j}^{-1} t_a x^{2k}}\right)\right]\right.\nonumber\\
&&\times\sum_{\vec n=0}^\infty\left[(-1)^{-\kappa\sum n_j-(N_f-\tilde N_f)\sum n_j/2}w^{\sum n_j}\left(\prod_{j=1}^{N}t_{b_j}^{\kappa n_j+(N_f-\tilde N_f)n_j/2}\right)\right.\nonumber\\
&&~~~~~~~~~~\tau^{\kappa\sum n_j-\tilde N_f\sum n_j}x^{\kappa\sum n_j^2+(N_f+\tilde N_f)\sum n_j/2+(N_f-\tilde N_f)\sum n_j^2/2}\nonumber\\
&&~~~~~~~~\left.\left(\prod_{i<j}^{N}t_{b_i}^{1/2}t_{b_j}^{-1/2}x^{n_i-n_j}-t_{b_i}^{-1/2}t_{b_j}^{1/2}x^{-(n_i-n_j)}\right)\left(\prod_{j=1}^{N}\prod_{k=0}^{n_j-1}\frac{\prod_{a=1}^{\tilde N_f}1-t_{b_j} \tilde t_a \tau^2 x^{2k}}{\prod_{a=1}^{N_f}1-t_{b_j} t_a^{-1}x^{2+2k}}\right)\right]\nonumber\\
&&\times\sum_{\vec{\bar n}=0}^\infty\left[(-1)^{\kappa\sum \bar n_j-(N_f-\tilde N_f)\sum\bar n_j/2}w^{-\sum\bar n_j}\left(\prod_{j=1}^{N}t_{b_j}^{-\kappa\bar n_j+(N_f-\tilde N_f)\bar n_j/2}\right)\right.\nonumber\\
&&~~~~~~~~~~\tau^{-\kappa\sum\bar n_j-\tilde N_f\sum\bar n_j}x^{-\kappa\sum\bar n_j^2+(N_f+\tilde N_f)\sum\bar n_j/2+(N_f-\tilde N_f)\sum\bar n_j^2/2}\nonumber\\
&&~~~~~~~~\left.\left.\left(\prod_{i<j}^{N}t_{b_i}^{1/2}t_{b_j}^{-1/2}x^{\bar n_i-\bar n_j}-t_{b_i}^{-1/2}t_{b_j}^{1/2}x^{-(\bar n_i-\bar n_j)}\right)\left(\prod_{j=1}^{N}\prod_{k=0}^{\bar n_j-1}\frac{\prod_{a=1}^{\tilde N_f}1-t_{b_j} \tilde t_a \tau^2 x^{2k}}{\prod_{a=1}^{N_f}1-t_{b_j} t_a^{-1} x^{2+2k}}\right)\right]\right\}.
\end{eqnarray}
More concisely,
\begin{eqnarray}\label{expression1}
&&I^{N_f>\tilde N_f}(x=e^{-\gamma},t=e^{iM},\tilde t=e^{i\tilde M},\tau=e^{i\mu},w,\kappa)\nonumber\\
&=&\frac{(-1)^{N(N-1)/2}}{N!}\sum_{b_1,\cdots,b_{N}=1}^{N_f}\left\{\left[\prod_{j=1}^{N}\prod_{k=0}^\infty\left(\frac{\prod_{a=1(\neq b_j)}^{N_f}1-t_{b_j} t_a^{-1}x^{2+2k}}{\prod_{a=1}^{\tilde N_f}1-t_{b_j} \tilde t_a \tau^2 x^{2k}}\right)\left(\frac{\prod_{a=1}^{\tilde N_f}1-t_{b_j}^{-1} \tilde t_a^{-1} \tau^{-2} x^{2+2k}}{\prod_{a=1(\neq b_j)}^{N_f}1-t_{b_j}^{-1} t_a x^{2k}}\right)\right]\right.\nonumber\\
&&\times\sum_{\vec n=0}^\infty\left[(-1)^{-\kappa\sum n_j-(N_f-\tilde N_f)\sum n_j/2}w^{\sum n_j}\left(\prod_{j=1}^{N}t_{b_j}^{\kappa n_j}\right)\tau^{\kappa\sum n_j}x^{\kappa\sum n_j^2}\right.\nonumber\\
&&~~~~~~~~\left.\left(\prod_{i<j}^{N}2\sinh\frac{iM_{b_i}-iM_{b_j}-2\gamma(n_i-n_j)}{2}\right)\left(\prod_{j=1}^{N}\prod_{k=0}^{n_j-1}\frac{\prod_{a=1}^{\tilde N_f}2\sinh\frac{-iM_{b_j}-i\tilde M_a-2i\mu+2\gamma k}{2}}{\prod_{a=1}^{N_f}2\sinh\frac{-iM_{b_j}+iM_a+2\gamma(1+k)}{2}}\right)\right]\nonumber\\
&&\times\sum_{\vec{\bar n}=0}^\infty\left[(-1)^{\kappa\sum \bar n_j-(N_f-\tilde N_f)\sum\bar n_j/2}w^{-\sum\bar n_j}\left(\prod_{j=1}^{N}t_{b_j}^{-\kappa\bar n_j}\right)\tau^{-\kappa\sum\bar n_j}x^{-\kappa\sum\bar n_j^2}\right.\nonumber\\
&&~~~~~~~~\left.\left.\left(\prod_{i<j}^{N}2\sinh\frac{iM_{b_i}-iM_{b_j}-2\gamma(\bar n_i-\bar n_j)}{2}\right)\left(\prod_{j=1}^{N}\prod_{k=0}^{\bar n_j-1}\frac{\prod_{a=1}^{\tilde N_f}2\sinh\frac{-iM_{b_j}-i\tilde M_a-2i\mu+2\gamma k}{2}}{\prod_{a=1}^{N_f}2\sinh\frac{-iM_{b_j}+iM_a+2\gamma(1+k)}{2}}\right)\right]\right\}\nonumber\\
\end{eqnarray}
where we identified some parameters as follows: $x= e^{-\gamma}$, $t_a= e^{iM_a}$, $\tilde t_a= e^{i\tilde M_a}$ and $\tau= e^{i\mu}$. If $b_i=b_j$ for $i\neq j$ the index vanishes because it has antisymmetric contributions of $n_i$ and $n_j$. Together with the flavor symmetry it implies that one can arrange $b_j$ in ascending order: $b_1<\cdots<b_{N}$ and thus replace $\frac{1}{N!}\sum_{b_1,\cdots,b_{N}=1}^{N_f}$ by $\sum_{1\leq b_1<\cdots<b_{N}\leq N_f}$. With some tricks described in the next section the index can be written as
\begin{eqnarray}\label{expression2}
&&I^{N_f>\tilde N_f}(x=e^{-\gamma},t=e^{iM},\tilde t=e^{i\tilde M},\tau=e^{i\mu},w,\kappa)\nonumber\\
&=&\sum_{1\leq b_1<\cdots<b_{N}\leq N_f}\nonumber\\
&&\textstyle\left\{\left(\prod_{\substack{i,j=1\\(i\neq j)}}^{N}2\sinh\frac{iM_{b_i}-iM_{b_j}}{2}\right)\left[\prod_{j=1}^{N}\prod_{k=0}^\infty\left(\frac{\prod_{a=1(\neq b_j)}^{N_f}1-t_{b_j} t_a^{-1}x^{2+2k}}{\prod_{a=1}^{\tilde N_f}1-t_{b_j} \tilde t_a \tau^2 x^{2k}}\right)\left(\frac{\prod_{a=1}^{\tilde N_f}1-t_{b_j}^{-1} \tilde t_a^{-1} \tau^{-2} x^{2+2k}}{\prod_{a=1(\neq b_j)}^{N_f}1-t_{b_j}^{-1} t_a x^{2k}}\right)\right]\right.\nonumber\\
&&\times\sum_{\vec n=0}^\infty\left[(-1)^{-\kappa\sum n_j-(N_f-\tilde N_f)\sum n_j/2}(-w)^{\sum n_j}\left(\prod_{j=1}^{N}t_{b_j}^{\kappa n_j}\right)\tau^{\kappa\sum n_j}x^{\kappa\sum n_j^2}\right.\nonumber\\
&&\textstyle~~~~~~~~\left.\left(\prod_{j=1}^{N}\prod_{k=0}^{n_j-1}\frac{\prod_{a=1}^{\tilde N_f}2\sinh\frac{-i\tilde M_a-iM_{b_j}-2i\mu+2\gamma k}{2}}{\left(\prod_{i=1}^{N}2\sinh\frac{iM_{b_i}-iM_{b_j}+2\gamma(k-n_i)}{2}\right)\left(\prod_{a=1(\notin\{b_j\})}^{N_f}2\sinh\frac{iM_a-iM_{b_j}+2\gamma(1+k)}{2}\right)}\right)\right]\nonumber\\
&&\times\sum_{\vec n=0}^\infty\left[(-1)^{\kappa\sum \bar n_j-(N_f-\tilde N_f)\sum\bar n_j/2}(-w)^{-\sum\bar n_j}\left(\prod_{j=1}^{N}t_{b_j}^{-\kappa\bar n_j}\right)\tau^{-\kappa\sum\bar n_j}x^{-\kappa\sum\bar n_j^2}\right.\nonumber\\
&&\textstyle~~~~~~~~\left.\left.\left(\prod_{j=1}^{N}\prod_{k=0}^{\bar n_j-1}\frac{\prod_{a=1}^{\tilde N_f}2\sinh\frac{-i\tilde M_a-iM_{b_j}-2i\mu+2\gamma k}{2}}{\left(\prod_{i=1}^{N}2\sinh\frac{iM_{b_i}-iM_{b_j}+2\gamma(k-\bar n_i)}{2}\right)\left(\prod_{a=1(\notin\{b_j\})}^{N_f}2\sinh\frac{iM_a-iM_{b_j}+2\gamma(1+k)}{2}\right)}\right)\right]\right\}.\qquad\qquad
\end{eqnarray}
Each of two summations corresponds to the $\mathcal N=2$ vortex and antivortex partition functions on $\mathbb R^2\times S^1$ respectively. Note that $(-1)^{(\ldots)}$ always give rise to a real valued factor because $\kappa+N_f/2+\tilde N_f$ is an integer.

The index for $N_f<\tilde N_f$ is simply obtained by interchanging $t_a\leftrightarrow\tilde t_a$ and $\kappa\rightarrow-\kappa$:
\begin{eqnarray}
&&I^{N_f<\tilde N_f}(x=e^{-\gamma},t=e^{iM},\tilde t=e^{i\tilde M},\tau=e^{i\mu},w,\kappa)\nonumber\\
&=&\sum_{1\leq b_1<\cdots<b_{N}\leq\tilde N_f}\nonumber\\
&&\textstyle\left\{\left(\prod_{\substack{i,j=1\\(i\neq j)}}^{N}2\sinh\frac{i\tilde M_{b_i}-i\tilde M_{b_j}}{2}\right)\left[\prod_{j=1}^{N}\prod_{k=0}^\infty\left(\frac{\prod_{a=1(\neq b_j)}^{\tilde N_f}1-\tilde t_{b_j}\tilde t_a^{-1}x^{2+2k}}{\prod_{a=1}^{N_f}1-\tilde t_{b_j}t_a \tau^2 x^{2k}}\right)\left(\frac{\prod_{a=1}^{N_f}1-\tilde t_{b_j}^{-1}t_a^{-1} \tau^{-2} x^{2+2k}}{\prod_{a=1(\neq b_j)}^{\tilde N_f}1-\tilde t_{b_j}^{-1}\tilde t_a x^{2k}}\right)\right]\right.\nonumber\\
&&\times\sum_{\vec n=0}^\infty\left[(-1)^{\kappa\sum n_j-(\tilde N_f-N_f)\sum n_j/2}(-w)^{\sum n_j}\left(\prod_{j=1}^{N}\tilde t_{b_j}^{-\kappa n_j}\right)\tau^{-\kappa\sum n_j}x^{-\kappa\sum n_j^2}\right.\nonumber\\
&&\textstyle~~~~~~~~\left.\left(\prod_{j=1}^{N}\prod_{k=0}^{n_j-1}\frac{\prod_{a=1}^{N_f}2\sinh\frac{-iM_a-i\tilde M_{b_j}-2i\mu+2\gamma k}{2}}{\left(\prod_{i=1}^{N}2\sinh\frac{i\tilde M_{b_i}-i\tilde M_{b_j}+2\gamma(k-n_i)}{2}\right)\left(\prod_{a=1(\notin\{b_j\})}^{\tilde N_f}2\sinh\frac{i\tilde M_a-i\tilde M_{b_j}+2\gamma(1+k)}{2}\right)}\right)\right]\nonumber\\
&&\times\sum_{\vec n=0}^\infty\left[(-1)^{-\kappa\sum \bar n_j-(\tilde N_f-N_f)\sum\bar n_j/2}(-w)^{-\sum\bar n_j}\left(\prod_{j=1}^{N}t_{b_j}^{\kappa\bar n_j}\right)\tau^{\kappa\sum\bar n_j}x^{\kappa\sum\bar n_j^2}\right.\nonumber\\
&&\textstyle~~~~~~~~\left.\left.\left(\prod_{j=1}^{N}\prod_{k=0}^{\bar n_j-1}\frac{\prod_{a=1}^{N_f}2\sinh\frac{-iM_a-i\tilde M_{b_j}-2i\mu+2\gamma k}{2}}{\left(\prod_{i=1}^{N}2\sinh\frac{i\tilde M_{b_i}-i\tilde M_{b_j}+2\gamma(k-\bar n_i)}{2}\right)\left(\prod_{a=1(\notin\{b_j\})}^{\tilde N_f}2\sinh\frac{i\tilde M_a-i\tilde M_{b_j}+2\gamma(1+k)}{2}\right)}\right)\right]\right\}.\qquad\qquad
\end{eqnarray}
For $N_f=\tilde N_f$ the integrand also has poles either at the origin or at the infinity depending on the sign of $N + \kappa m_j$. For $N + \kappa m_j > 0$ the residue at the origin is given by
\begin{eqnarray}
&&\textrm{Res}(\ldots,0)\nonumber\\
&=&x^{-\sum_{i\neq j}|m_i-m_j|/2+N_f\sum|m_j|}\tau^{-N_f\sum|m_j|}\left(\prod_{j=1}^{N}\lim_{z_j\rightarrow0}\frac{1}{(N+\kappa m_j - 1)!}\frac{\partial^{N + \kappa m_j - 1}}{\partial z_j^{N + \kappa m_j - 1}}\right)\nonumber\\
&&\left[\prod_{j=1}^{N}\left(\prod_{i=1(\neq j)}^{N}z_j-z_ix^{|m_i-m_j|}\right)\left(\prod_{a=1}^{N_f}\prod_{k=0}^\infty\frac{z_j-t_a^{-1}\tau^{-1}x^{|m_j|+2+2k}}{1-z_j t_a\tau x^{|m_j|+2k}}\frac{1-z_j \tilde t_a^{-1}\tau^{-1}x^{|m_j|+2+2k}}{z_j-\tilde t_a\tau x^{|m_j|+2k}}\right)\right]\nonumber\\
\end{eqnarray}
Let us first consider the $N=1,\kappa = 0$ case. In that case one has a vanishing infinite product:
\begin{equation}
\sim\prod_{k=0}^{\infty}t_a^{-1}\tilde t_a^{-1}\tau^{-2}x^2\rightarrow0
\end{equation}
assuming $|t_a^{-1}\tilde t_a^{-1}\tau^{-2}x^2|<1$, which doesn't spoil the original range of parameters that we already assumed at start. For general $N$ and $\kappa$, there are $N + \kappa m_j - 1$ differentiations. When each of them acts on the above infinite product, an additional factor arises. Nevertheless, one still has a vanishing infinite product because there are only the finite number of such additional factors, which are not singular. In the same manner the residue at the infinity also vanishes. Therefore, since the residues at the other poles are the same as those for $N_f\neq\tilde N_f$, both results for $N_f>\tilde N_f$ and $N_f<\tilde N_f$ even work for $N_f=\tilde N_f$.

One can  rewrite the factorized index using the permutation of the chemical potentials $t_{a}$:
\be
&&I^{N_f,\tilde{N}_f}(x,t,\tilde{t},w,\kappa) \nonumber \\
&\!=\!&\!\! \frac{1}{N!(N_f\!-\!N)!}\!\sum_{\sigma(t)}I^{pert}(x,\sigma(t),\tilde{t},\tau) \!\!\!\left[\sum_{\vec{n}=0}^\infty(-w)^{n}I_{\{n_j\}}(x,\sigma(t),\tilde{t},\tau,\kappa)\!\right]
    \!\!\!\left[\sum_{\vec{\bar{n}}=0}^\infty(-w)^{-\bar{n}}I_{\{\bar{n}_j\}}(x,\sigma(t),\tilde{t},\tau,-\kappa)\!\right]\nonumber\\
\ee
where $n=\sum_j n_j,\bar{n}=\sum_j\bar{n}_j$ and $\sigma(t)$
denotes the permutation of $t_a$'s. Here the perturbative and vortex
contributions are given by \be
I^{pert}(x,t,\tilde{t},\tau) \!&\!=\!&\! \prod_{i\neq j}^N2\sinh\frac{iM_i-iM_j}{2}
    \prod_{j=1}^N\prod_{k=0}^\infty\!\!\left[\!\prod_{a=1(\neq j)}^{N_f}\frac{1\!-\!t_j t_a^{-1}x^{2k+2}}{1\!-\!t^{-1}_j t_a x^{2k}}\right]\!\!\!\!\left[\prod_{a=1}^{\tilde{N}_f}\frac{1\!-\!t_j^{-1}\tilde{t}_a^{-1}\tau^{-2}x^{2k+2}}{1\!-\!t_j\tilde{t}_a
    \tau^2x^{2k}}\right], \nn \\
I_{\{k_j\}}(x,t,\tilde{t},\tau,\kappa)\! &\!=\!&\! (-1)^{-\kappa n-(N_f-\tilde N_f)n/2}e^{i\kappa\sum_j( M_j n_j+\mu n_j+i\gamma n_j^2)}\nonumber\\
&&\prod_{j=1}^N\prod_{k=1}^{n_j}\frac{\prod_{a=1}^{\tilde{N}_f}2\sinh\frac{-i\tilde{M}_a-iM_j-2i\mu+2\gamma (k-1)}{2}}
    {\prod_{i=1}^N2\sinh\frac{iM_i-iM_j+2\gamma(k-1-n_i)}{2}\prod_{a=N+1}^{N_f}2\sinh\frac{iM_a-iM_j+2\gamma k}{2}}\qquad .
\ee
One can compare the vortex partition function obtained here with the $\mathcal N=4$ result in \cite{KKKL12}. In order to compare our result to that of \cite{KKKL12} one would set $N_f=\tilde N_f$, restore the $R$-charge to $\frac{1}{2}$ and restrict $b_j=j\in\{1,\cdots,N\}$; for the result in \cite{KKKL12}, set $\gamma'=0$, $R=-\tilde R=\frac{1}{2}$ and redefine $2i\gamma\rightarrow\gamma$. A factor $\left(\prod_{i=1}^{N}\sinh\frac{iM_{b_i}-iM_{b_j}+2\gamma(k-n_i)}{2}\right)^{-1}$ corresponds to $-z_v z_{fund}^{N}$ in \cite{KKKL12}; likewise,
\begin{eqnarray}
&&\prod_{a=1(\notin\{b_j\})}^{N_f}\frac{1}{\sinh\frac{iM_a-iM_{b_j}+2\gamma(1+k)}{2}}=-z_{fund}^{N_f-N},\\
&&\prod_{a=1}^{N_f}\sinh\frac{-i\tilde M_a-iM_{b_j}-2i\mu+2\gamma (\frac{1}{2}+k)}{2}=-z_{anti}^{N_f}
\end{eqnarray}
Mass parameters of each result are identified as follows:
\begin{equation}
iM_j+i\mu=\mu_j+2\gamma+\delta,\qquad iM_a+i\mu=\mu_a+\delta,\qquad i\tilde M_b+i\mu=-\mu_b-\delta
\end{equation}
where $j=0,\cdots,N$; $a=N+1,\cdots,N_f$; $b=1,\cdots,N_f$ and $\delta$ is an undetermined parameter coming from the gauge symmetry. Asymmetry between $iM_j$ and $iM_a$ might come from the fact that $N$ flavors have nonzero VEVs while $N_f-N$ flavors do not.

\section{Detailed calculations}

In this section we will see an identity that can be used when one derives \eqref{expression2} from \eqref{expression1}. First let us focus on $\prod_{j=1}^{N}\prod_{k=0}^{n_j-1}\prod_{a=1}^{N_f}2\sinh\frac{-iM_{b_j}+iM_a+2\gamma(1+k)}{2}$ in the denominator of the vortex partition part. It decomposes as
\begin{eqnarray}
&&\prod_{j=1}^{N}\prod_{k=0}^{n_j-1}\prod_{a=1}^{N_f}2\sinh\frac{-iM_{b_j}+iM_a+2\gamma(1+k)}{2}\nonumber\\
&=&\prod_{j=1}^{N}\prod_{k=0}^{n_j-1}\left(\prod_{i=1}^{N}2\sinh\frac{iM_{b_i}-iM_{b_j}+2\gamma(1+k)}{2}\right)\left(\prod_{a=1(\notin\{b_j\}}^{N_f}2\sinh\frac{iM_a-iM_{b_j}+2\gamma(1+k)}{2}\right).\nonumber\\
\end{eqnarray}
The former factor can be written as
\begin{eqnarray}
&&\prod_{i,j=1}^{N}\prod_{k=0}^{n_j-1}2\sinh\frac{iM_{b_i}-iM_{b_j}+2\gamma(1+k)}{2}\nonumber\\
&=&\left(\prod_{j=1}^{N}\prod_{k=0}^{n_j-1}2\sinh\gamma(1+k)\right)\nonumber\\
&&\left[\prod_{i<j}^{N}\left(\prod_{k=0}^{n_j-1}2\sinh\frac{iM_{b_i}-iM_{b_j}+2\gamma(1+k)}{2}\right)\left(\prod_{k=0}^{n_i-1}2\sinh\frac{iM_{b_j}-iM_{b_i}+2\gamma(1+k)}{2}\right)\right]\nonumber\\
&=&\left(\prod_{j=1}^{N}\prod_{k=0}^{n_j-1}2\sinh\gamma(1+k)\right)\nonumber\\
&&\left[\prod_{i<j}^{N}(-1)^{n_i}\left(\prod_{k=0}^{n_j-1}2\sinh\frac{iM_{b_i}-iM_{b_j}+2\gamma(1+k)}{2}\right)\left(\prod_{k=0}^{n_i-1}2\sinh\frac{iM_{b_i}-iM_{b_j}+2\gamma(k-n_i)}{2}\right)\right]\nonumber\\
&=&\left(\prod_{j=1}^{N}\prod_{k=0}^{n_j-1}2\sinh\gamma(1+k)\right)\nonumber\\
&&\left(\prod_{i<j}^{N}2\sinh\frac{iM_{b_i}-iM_{b_j}}{2}\right)^{-1}\left(\prod_{i<j}^{N}(-1)^{n_i}\prod_{k=0}^{n_i+n_j}2\sinh\frac{iM_{b_i}-iM_{b_j}+2\gamma(k-n_i)}{2}\right).
\end{eqnarray}
Also note
\begin{eqnarray}
\hspace{-1cm}&&\prod_{i,j=1}^{N}\prod_{k=0}^{n_j-1}2\sinh\frac{iM_{b_i}-iM_{b_j}+2\gamma(k-n_i)}{2}\nonumber\\
\hspace{-1cm}&=&\left(\prod_{j=1}^{N}\prod_{k=0}^{n_j-1}2\sinh\gamma(k-n_j)\right)\nonumber\\
\hspace{-1cm}&&\left[\prod_{i<j}^{N}\left(\prod_{k=0}^{n_j-1}2\sinh\frac{iM_{b_i}-iM_{b_j}+2\gamma(k-n_i)}{2}\right)\left(\prod_{k=0}^{n_i-1}2\sinh\frac{iM_{b_j}-iM_{b_i}+2\gamma(k-n_j)}{2}\right)\right]\nonumber\\
\hspace{-1cm}&=&(-1)^{\sum n_j}\left(\prod_{j=1}^{N}\prod_{k=0}^{n_j-1}2\sinh\gamma(1+k)\right)\nonumber\\
\hspace{-1cm}&&\left[\prod_{i<j}^{N}(-1)^{n_i}\left(\prod_{k=0}^{n_j-1}2\sinh\frac{iM_{b_i}-iM_{b_j}+2\gamma(k-n_i)}{2}\right)\left(\prod_{k=0}^{n_i-1}2\sinh\frac{iM_{b_i}-iM_{b_j}+2\gamma(n_j-n_i+1+k)}{2}\right)\right]\nonumber\\
\hspace{-1cm}&=&(-1)^{\sum n_j}\left(\prod_{j=1}^{N}\prod_{k=0}^{n_j-1}2\sinh\gamma(1+k)\right)\nonumber\\
\hspace{-1cm}&&\left(\prod_{i<j}^{N}2\sinh\frac{iM_{b_i}-iM_{b_j}-2\gamma(n_i-n_j)}{2}\right)^{-1}\left(\prod_{i<j}^{N}(-1)^{n_i}\prod_{k=0}^{n_i+n_j}2\sinh\frac{iM_{b_i}-iM_{b_j}+2\gamma(k-n_i)}{2}\right).
\end{eqnarray}
Combining those results one obtains the following identity:
\begin{eqnarray}
\hspace{-1cm}&&\sum_{\vec n=0}^\infty\left[w^{\sum n_j}\left(\prod_{i<j}^{N}2\sinh\frac{iM_{b_i}-iM_{b_j}-2\gamma(n_i-n_j)}{2}\right)\left(\prod_{j=1}^{N}\prod_{k=0}^{n_j-1}\frac{\prod_{a=1}^{\tilde N_f}2\sinh\frac{-iM_{b_j}-i\tilde M_a-2i\mu+2\gamma k}{2}}{\prod_{a=1}^{N_f}2\sinh\frac{-iM_{b_j}+iM_a+2\gamma(1+k)}{2}}\right)\right]\nonumber\\
\hspace{-1cm}&=&\left(\prod_{i<j}^{N}2\sinh\frac{iM_{b_i}-iM_{b_j}}{2}\right)\nonumber\\
\hspace{-1cm}&&\times\sum_{\vec n=0}^\infty\left[(-w)^{\sum n_j}\left(\prod_{j=1}^{N}\prod_{k=0}^{n_j-1}\frac{\prod_{a=1}^{\tilde N_f}2\sinh\frac{-i\tilde M_a-iM_{b_j}-2i\mu+2\gamma k}{2}}{\left(\prod_{i=1}^{N}2\sinh\frac{iM_{b_i}-iM_{b_j}+2\gamma(k-n_i)}{2}\right)\left(\prod_{a=1(\notin\{b_j\})}^{N_f}2\sinh\frac{iM_a-iM_{b_j}+2\gamma(1+k)}{2}\right)}\right)\right].\nonumber\\
\hspace{-1cm}
\end{eqnarray}

\end{document}